\documentclass[a4paper,12pt]{article}
\usepackage[top=3cm,bottom=2cm,left=3cm,right=3cm]{geometry}

\usepackage{amsmath}
\usepackage{graphicx}
\usepackage{bm}
\usepackage{lineno}
\usepackage{subfigure}
\usepackage{url}
\usepackage{authblk}
\usepackage[colorlinks,unicode=true]{hyperref}
\usepackage{ulem}
\newcounter{firstbib}

\begin{document}

\title{Detailed polarization measurements of the prompt emission of five Gamma-Ray Bursts}
\author[]{Shuang-Nan Zhang\thanks{Corresponding Author. These authors contributed equally:  
Shuang-Nan Zhang, Merlin Kole. }}
\author[]{Merlin Kole$^\ast$}
\author[]{Tian-Wei Bao}
\author[]{Tadeusz Batsch}
\author[]{Tancredi Bernasconi}
\author[]{Franck Cadoux}
\author[]{Jun-Ying Chai}
\author[]{Zi-Gao Dai}
\author[]{Yong-Wei Dong}
\author[]{Neal Gauvin}
\author[]{Wojtek Hajdas}
\author[]{Mi-Xiang Lan}
\author[]{Han-Cheng Li}
\author[]{Lu Li}
\author[]{Zheng-Heng Li}
\author[]{Jiang-Tao Liu}
\author[]{Xin Liu}
\author[]{Radoslaw Marcinkowski}
\author[]{Nicolas Produit}
\author[]{Silvio Orsi}
\author[]{Martin Pohl}
\author[]{Dominik Rybka}
\author[]{Hao-Li Shi}
\author[]{Li-Ming Song}
\author[]{Jian-Chao Sun}
\author[]{Jacek Szabelski}
\author[]{Teresa Tymieniecka}
\author[]{Rui-Jie Wang}
\author[]{Yuan-Hao Wang}
\author[]{Xing Wen}
\author[]{Bo-Bing Wu}
\author[]{Xin Wu}
\author[]{Xue-Feng Wu}
\author[]{Hua-Lin Xiao}
\author[]{Shao-Lin Xiong}
\author[]{Lai-Yu Zhang}
\author[]{Li Zhang}
\author[]{Xiao-Feng Zhang}
\author[]{Yong-Jie Zhang}
\author[]{Anna Zwolinska}

\affil[]{}

\maketitle



{\bf Gamma-ray bursts are the strongest explosions in the Universe since the Big Bang, believed to be produced either in forming black holes at the end of massive star evolution ~\cite{Woosley1993,Iwamoto,MacFayden1999} or merging of compact objects~\cite{Gehrels2013}. Spectral and timing properties of gamma-ray bursts suggest that the observed bright gamma-rays are produced in the most relativistic jets in the Universe~\cite{Gehrels2013}; however, the physical properties, especially the structure and magnetic topologies in the jets are still not well known, despite several decades of studies. It is widely believed that precise measurements of the polarization properties of gamma-ray bursts should provide crucial information on the highly relativistic jets~\cite{Toma2009}. As a result there have been many reports of gamma-ray burst polarization measurements with diverse results, see~\cite{Covino:2016cuw}, however many such measurements suffered from substantial uncertainties, mostly systematic ~\cite[and references therein]{MCCONNELL20171}. After the first  successful measurements by the GAP and COSI instruments ~\cite{COSI2,GAPGRB2,GAP}, here we report a statistically meaningful sample of precise polarization measurements, obtained with the dedicated gamma-ray burst polarimeter, POLAR onboard China's Tiangong-2 spacelab. Our results suggest that the gamma-ray emission is at most polarized at a level lower than some popular models have predicted; although our results also show intrapulse evolution of the polarization angle. This indicates that the low polarization degrees could be due to an evolving polarization angle during a gamma-ray burst.}


POLAR is a dedicated Gamma-ray Burst (GRB) polarization detection experiment onboard China's Tiangong-2 spacelab~\cite{Produit2017}, launched on Sept. 15th, 2016 and stopped operation on March 31, 2017. POLAR detected 55 GRBs with high significance. In order to make statistically significant GRB polarization measurements and yet with negligible systematic errors, we select a subsample of five GRBs for detailed analysis of their polarization properties; please refer to the supplementary information (SI) for the sample selection criteria and the properties of the five selected GRBs. We employ a straight forward $\chi^2$ based analysis, similar to that successfully employed in~\cite{GAP}, to study the polarization properties of the five GRBs, while a Bayesian method is employed to accurately determine the credible regions of the measurements. The studies rely on extensive ground and in-orbit calibration data and Monte-Carlo simulations matching the calibration data~\cite{LI2018,Kole2017}. Please refer to the methods section for details of the methodology and analysis.

\begin{figure}[!ht]
   \centering
     \includegraphics[width=15 cm]{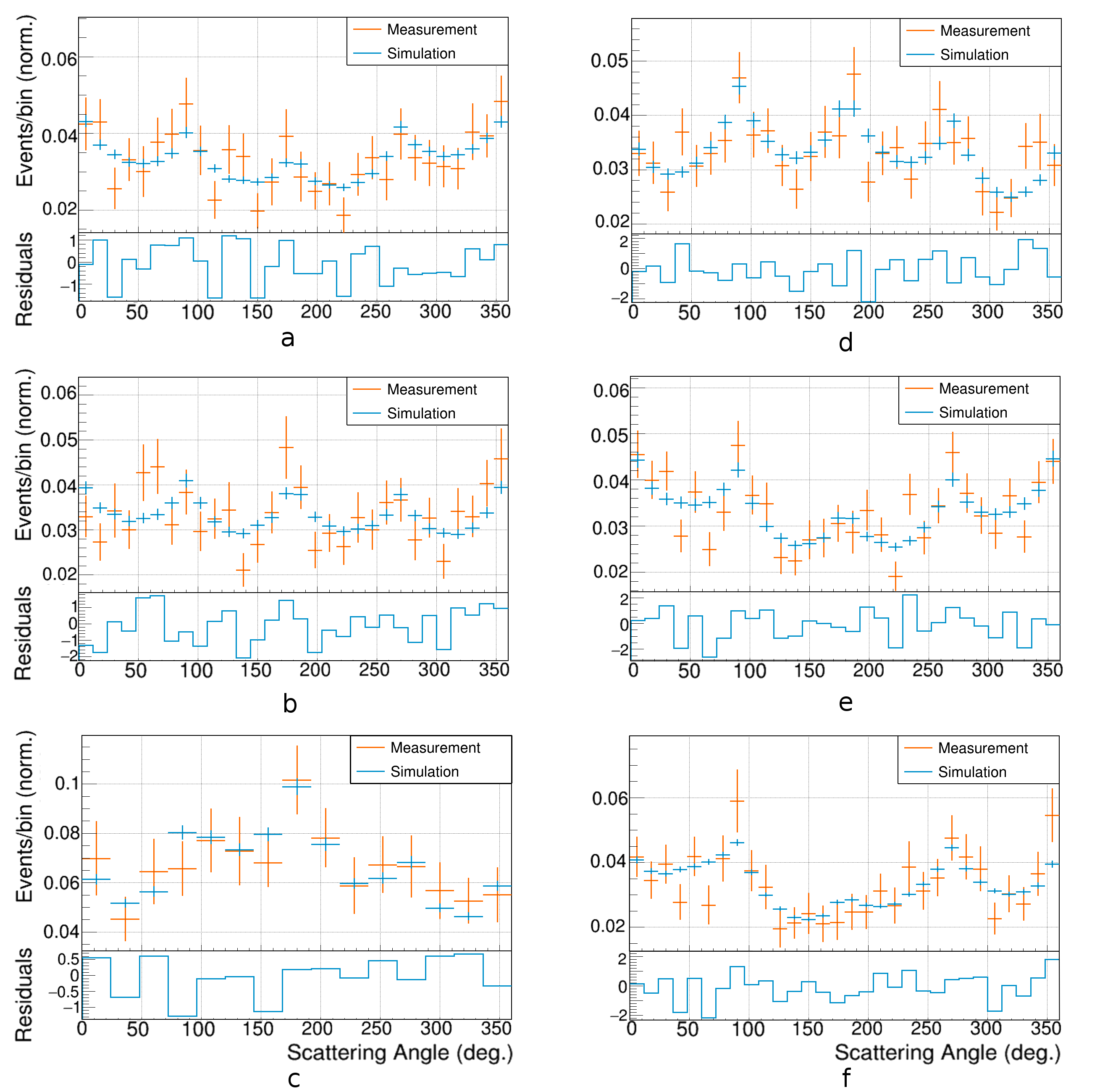}
   \caption[Modulation curves of the 5 GRBs]
  {\textbf{Measured and best fitting simulated modulation curves for all 5 GRBs and the second time interval of GRB 170114A.} Orange crosses are the measured modulation curves normalized to have a total bin content of unity, after subtracting background. The uncertainties in the orange histograms contain the statistical uncertainties. A grid of 6060 equally normalized simulated modulation curves was produced with different values for the polarization degree and angle. The blue crosses are the Monte Carlo produced modulation curves best fitting the measured data, the uncertainties displayed here contain both the statistical and systematic uncertainty. The blue histograms below are the fitting residuals. The panels on the left from top to bottom are 161218A (a), 170101A (b) and 170127C (c) and on the  right from top to bottom are 170206A (d), 170114A (e) and the second half (3 seconds) of 170114A (f). The detailed fitting results are listed in Table~\ref{tab:res_summary}. \\}
 \label{fig:mod_comb}
 \end{figure}

In Figure~\ref{fig:mod_comb}, we show the measured modulation curves of the five GRBs integrated over the whole GRB duration, together with the best fitting simulated modulation curves from linear polarization and fitting residuals. All fittings are statistically acceptable with no significant systematic deviations. In Figure~\ref{fig:comb_post}, we show the 2-D posterior distributions of the five GRBs, i.e., the posterior probability as functions of both polarization angle (PA) and degree (PD). Clearly the measured PD is correlated with PA for all five GRBs; we thus calculate the cumulative probabilities of PD marginalized over the PA for all five GRBs, as shown in Figure~\ref{fig:comb_CDF}. In Table~\ref{tab:res_summary} we summarize the main results for the five GRBs, including for each GRB the name, T90 value~\cite{1995AAS...186.5301K} (the time interval over which $90\%$ of the total background-subtracted counts are observed), the fluence, the most likely value of PD, the compatibility of the time integrated polarization with an unpolarized flux, the upper limit of the PD, and the most likely PA and evidence for a change in the PA.

We conclude that for the PD measurement, the most probable PD values for all five GRBs are between 4\% to 11\%. The 99\% upper limit ranges between 28\% to 45\% for four GRBs and 67\% for GRB 170127C, which has the lowest fluence and thus the poorest statistics for polarization measurement. Although the analysis provides non-zero polarization degrees, it should be noted that linear polarization measurements always provide positive results, see for example~\cite[and references therein]{Maier2014} for a discussion. The low polarization degrees found in the analysis presented here do not allow us to fully reject the hypothesis that every GRB is non-polarized. However, from Figure~\ref{fig:comb_CDF} we can find for each GRB the probability that PD is less than a certain positive value, which turns out to be between 5.8\% and 14\% for 2\% PD. Since all the five GRBs show statistically consistent PD, we can combine the measurements to give joint lower and upper limits. As can be found in the bottom right panel of Figure~\ref{fig:comb_CDF}, the average PD is 10\% and the probability is 0.1\% that the polarization degrees of all the five GRBs are lower than 5\% or higher than 16\%. Our results indicate for the first time with high precision that the prompt gamma-ray emission of GRBs is not highly polarized, as some models have predicted. Furthermore the results favor a low polarization level and can reject the hypothesis that all GRBs are unpolarized while several individual GRB polarization levels are still found to be consistent with zero polarization. Further evidence for a low but non-zero PD values was found when performing time dependent analysis.

Previously it has been reported that the PA of GRBs can change during the GRB~\cite{GAP}. We thus divide each of all five GRBs into two equal time bins and make the same analysis, to examine if the PA changed during each burst. As shown in detail in the methods section, a significant PA change was not observed for three GRBs, can not be constrained for GRB 170127C due to low statistics but was clearly detected for GRB 170114A. Interestingly, GRB 170114A is a single pulse GRB and also has the lowest PD (4\%) among all five GRBs integrated over the whole GRB periods. Performing a time resolved analysis, similar to that applied previously by the GAP collaboration on the long multi-peak GRB 100826A, we find that the polarization properties of this GRB are best described by a constant PD of $28\%$ with a PA which changes significantly during three 2 second time bins. While the time integrated analysis of this GRB is consistent with an unpolarized flux, the time resolved analysis results in a PD which is inconsistent with an unpolarized flux with a $99.7\%$ confidence level. While the measurement results of GAP for GRB 100826A already showed strong evidence of changes in the PA for different peaks inside of a long GRB, the measurement of GRB 170114A shows an intrapulse evolution of the PA. This implies that the measured low PD for all GRBs in the sample presented here integrated over the whole burst duration might be due to rapid changes of PA during the bursts. This however cannot be well constrained with the current instruments.

In summary, with the dedicated GRB polarization instrument POLAR onboard China's Tiangong-2 spacelab, we have measured the polarization properties of five GRBs with high precision, so far the largest sample of GRB polarization measurements with high statistical significance. For the first time we find that averaged over the whole bursts, GRBs are at most modestly polarized, i.e, PD$\approx$10\%. For the brightest GRB in the sample an unpolarized flux can be rejected using time resolved analysis and a PD of $28\pm9\%$ is found. Furthermore for the first time an intrapulse evolution of the PA is detected which has important implications for the time integrated PD measurements, as these low measurement values of the PD of all GRBs might be due to rapid PA changes during GRBs. Statistically most of previous prompt polarization emission measurements of GRBs, though some of quite high polarization degrees at face values, are consistent with our results, given their larger uncertainties in most cases. A summary of all previous reported GRB prompt emission polarization measurements is given in Supplementary Table 1 of the Supplementary Material.

Regarding the theoretical interpretation there are three main factors that affect the polarization properties significantly, these are the emission mechanism~\cite{KCWestfold1959,WHMcMaster1961,Lazzati2004}, the magnetic field configuration (MFC)~\cite[and references therein]{JG2003,Toma2009,MXLan2018}, and the jet geometry~\cite{Rossi2004,Gill2018}. Several popular models have been proposed to interpret GRBs, e.g., internal shock~\cite{MJRees1994,BP1994}, dissipative photosphere~\cite{PMeszaros2000,MJRees2005}, and internal-collision-induced magnetic reconnection and turbulence (ICMART)~\cite{BingZhang2011}. Depending on the MFC in the collision shell, the predicted polarization degree (PD) of the internal shock model can vary from a few\% to 70\%, about 10\% PD for the mixed MFC and up to about 70\% for the large-scale ordered MFC. For the dissipative photosphere model, the predicted PD is relatively low in the $\gamma$-ray band~\cite{Christoffer2016}. A decaying PD is predicted during each pulse in the ICMART model; the maximum value at the beginning of the pulse can reach up to 60\% and the minimum value at the end decreases to a few \%-10\% as presented in  ~\cite{BingZhang2011}. However, the detailed numerical simulations show that $\sim10\%$ PD is also possible for the ICMART model~\cite{WeiDeng2016}. As our results overall show a relatively low PD at first sight they appear to agree with the dissipative photosphere model. However, this model has difficulty to explain the PD of $28\%$ detected for GRB 170114A using the time binned analysis as well as the change in PA which appears to be of a continuous nature. The average PD level of 10\%  is possible according to the ICMART model, depending on the MFC in the emission region. For the internal shock model, the predicted PD can range from few\% to 70\%, depending on the MFC in the collision shells. The observed $\sim10\%$ PD with POLAR suggests that the MFC is most likely mixed during the GRB prompt phase. Alternatively, a violent change of the PA can also result in a decrease of the time-integrated PD, which should have significant implications to all GRB models. To further understand the physics and astrophysics of the most relativistic jets produced by the strongest explosions in the Universe, future GRB polarization instruments are required to provide time-resolved polarization properties of much larger samples of GRBs.


\renewcommand{\refname}{References}

\section*{}

The authors declare no competing financial interests. Correspondence and requests for materials should be addressed to Shuang-Nan Zhang (zhangsn@ihep.ac.cn) and/or Merlin Kole (Merlin.Kole@unige.ch).

\section*{Acknowledgments}

We gratefully acknowledge the financial support from the National Basic Research Program (973 Program) of China (Grant No. 2014CB845800), the Joint Research Fund in Astronomy under the cooperative agreement between the National Natural Science Foundation of China and the Chinese Academy of Sciences (Grant No. U1631242), the National Natural Science Foundation of China (Grant No. 11503028, 11403028), the Strategic Priority Research Program of the Chinese Academy of Sciences (Grant No. XDB23040400), the Swiss Space Office of the State Secretariat for Education, Research and Innovation (ESA PRODEX Programme), the National Science Center of Poland (Grant No. 2015/17/N/ST9/03556), and the Youth Innovation Promotion Association of Chinese Academy of Sciences (Grant No. 2014009). We furthermore thank Dr. J.M. Burgess of MPE, Garching, Germany, for providing the energy spectra for 170114A.

\section*{Authors Contributions}

The author list is in alphabetical order except the two first names, who are the first authors. T.W. Bao, T.Batsch, T. Bernasconi, F. Cadoux, J.Y. Chai, Y.W. Dong, N. Gauvin, W. Hajdas, M. Kole, H.C. Li, L. Li, Z.H. Li, J.T. Liu, X. Liu, R. Marcinkowski, S. Orsi, M. Pohl, N. Produit, D. Rybka, H.L. Shi, L.M. Song, J.C. Sun, J. Szabelski, T. Tymieniecka, R.J. Wang, X. Wen, B.B. Wu, X. Wu, H.L. Xiao, S.L. Xiong, L.Y. Zhang, L. Zhang, S.N. Zhang, X.F. Zhang, Y.J. Zhang and A. Zwolinska contributed to the development of the mission concept and/or construction and testing of POLAR. M. Kole, Z.H. Li, N. Produit, J.C. Sun, Y.H. Wang, S.L. Xiong and S.N. Zhang were involved in the presented analysis. Z.G. Dai, M.X. Lan and X.F. Wu contributed to the theoretical discussions in this manuscript. The manuscript was produced by M. Kole, Z.H. Li, J.C. Sun, Y.H. Wang and S.N. Zhang. The PIs of the POLAR collaboration are S.N. Zhang, M. Pohl and X. Wu.

\section*{Author Information}

\textit{Key Laboratory of Particle Astrophysics, Institute of High Energy Physics, Chinese Academy of Sciences, Beijing 100049, China}\newline
Shuang-Nan Zhang, Tian-Wei Bao, Jun-Ying Chai, Yong-Wei Dong, Han-Cheng Li, Lu Li, Zheng-Heng Li, Jiang-Tao Liu, Xin Liu, Hao-Li Shi, Li-Ming Song, Jian-Chao Sun, Rui-Jie Wang, Yuan-Hao Wang, Xing Wen, Bo-Bing Wu, Hua-Lin Xiao, Shao-Lin Xiong, Lai-Yu Zhang, Li Zhang, Xiao-Feng Zhang, Yong-Jie Zhang
\newline\newline
\textit{University of Chinese Academy of Sciences, Beijing 100049, China}\newline
Shuang-Nan Zhang, Jun-Ying Chai, Han-Cheng Li, Zheng-Heng Li, Xin Liu, Li-Ming Song, Yuan-Hao Wang, Xing Wen \newline\newline
\textit{Department of Nuclear and Particle Physics, University of Geneva, 24 Quai Ernest-Ansermet, 1205 Geneva, Switzerland}\newline
Merlin Kole, Franck Cadoux, Silvio Orsi, Martin Pohl, Xin Wu \newline\newline
\textit{National Centre for Nuclear Research, ul. A. Soltana 7, 05-400 Otwock, Swierk, Poland} \newline
Tadeusz Batsch, Dominik Rybka, Jacek Szabelski, Teresa Tymieniecka, Anna Zwolinska \newline\newline
\textit{University of Geneva, Geneva Observatory, ISDC, 16, Chemin d'Ecogia, 1290 Versoix, Switzerland} \newline
Tancredi Bernasconi, Neal Gauvin, Nicolas Produit \newline\newline
\textit{School of Astronomy and Space Science, Nanjing University, Nanjing 210093, China} \newline
Zi-Gao Dai, Mi-Xiang Lan \newline\newline
\textit{Key Laboratory of Modern Astronomy and Astrophysics (Nanjing University), Ministry of Education, China} \newline
Zi-Gao Dai \newline\newline
\textit{Paul Scherrer Institut, 5232, Villigen, Switzerland}\newline
Wojtek Hajdas, Radoslaw Marcinkowski, Hua-Lin Xiao \newline\newline
\textit{Purple Mountain Observatory, Chinese Academy of Sciences, Nanjing 210008, China} \newline
Mi-Xiang Lan, Xue-Feng Wu 

\begin{figure}[!ht]
   \centering
     \includegraphics[width=15 cm]{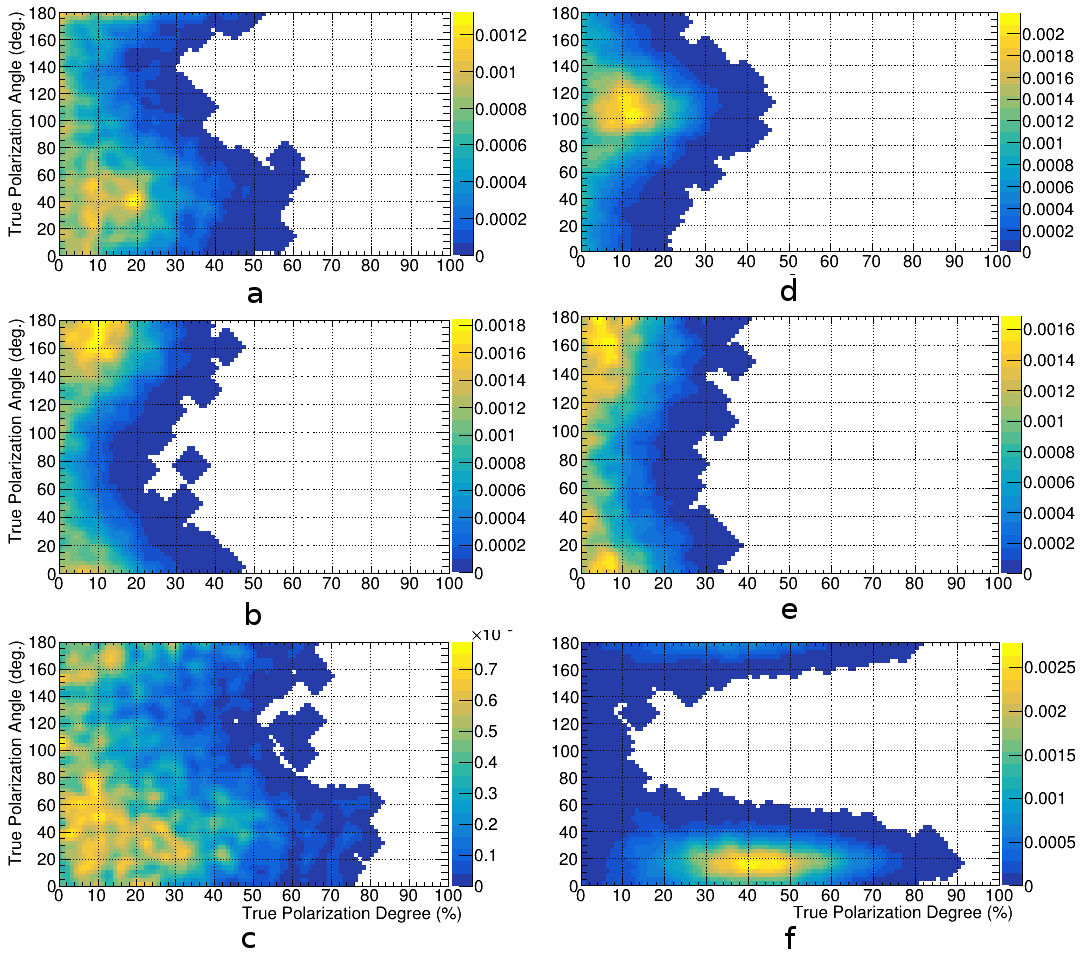}
   \caption[Posterior distributions of the 5 GRBs and the second time interval of GRB 170114A. ]
 {\textbf{Posterior distributions of the polarization parameters for the 5 GRBs and the second time intervals of GRB 170114A.} The panels on the left from top to bottom are 161218A (a), 170101A (b) and 170127C (c) and on the  right from top to bottom are 170206A (d), 170114A (e) and the second half (3 seconds) of 170114A (f).}
 \label{fig:comb_post}
 \end{figure}

\begin{figure}[!ht]
   \centering
     \includegraphics[width=15 cm]{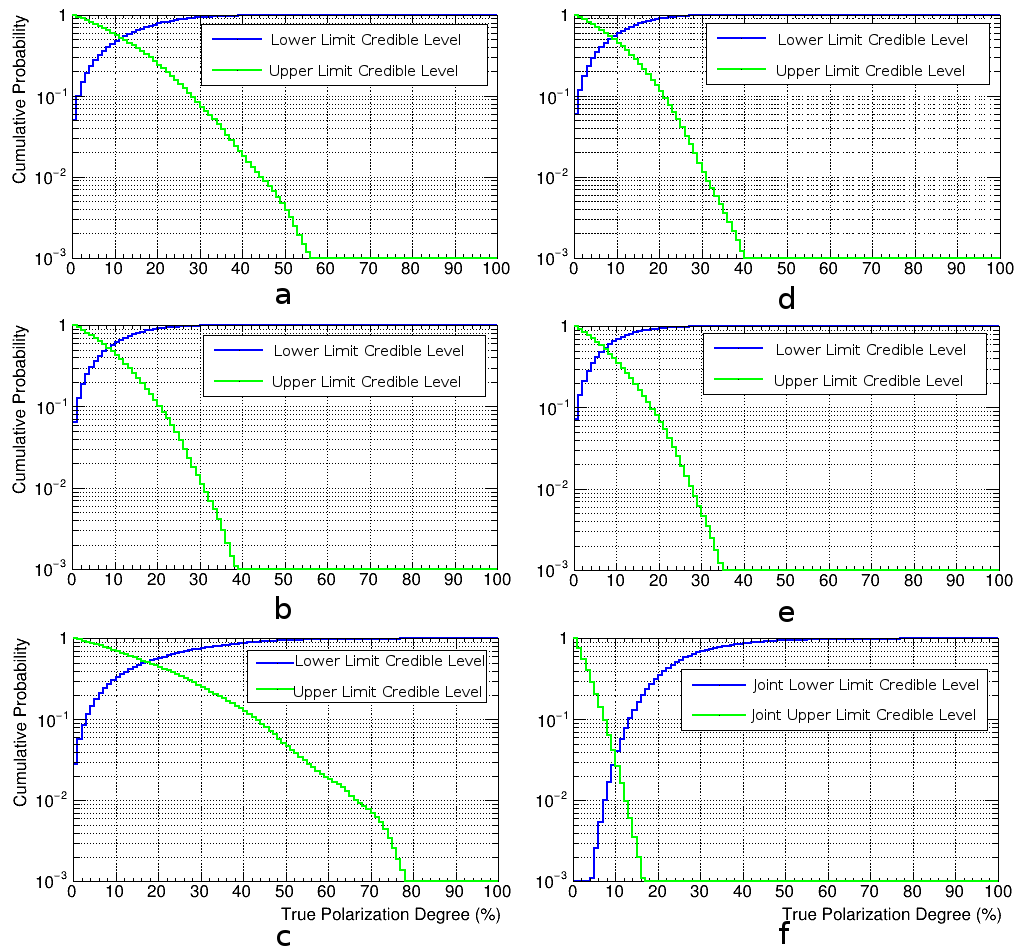}
   \caption[Cumulative probability functions for the 5 GRBs]
 {\textbf{Cumulative probability functions of the polarization degree for the studied GRB samples.} The lower (blue) and upper (green) credible levels of the polarization degree of the five GRBs are shown.  The panels on the left from top to bottom are 161218A (a), 170101A (b) and 170127C (c) and on the  right from top to bottom are 170206A (d), 170114A (e) and the joint lower and upper credible levels of the five GRBs (f). In each panel, the blue curve is the lower credible level: the cumulative probability that the polarization degree is below the value given by the curve. Similarly, the green curve is the upper credible level: the cumulative probability that the polarization degree is above the value given by the curve. Plotted in the bottom right panel are the credible levels for the polarization degrees of all the five GRBs to be lower or higher than the values given by the blue or green curves, respectively.}
 \label{fig:comb_CDF}
 \end{figure}

\begin{table}[!ht]
  \scriptsize
  \centering
 \caption{\textbf{Summary of the five GRBs.} The different properties of the 5 GRBs and of 2 time bins of GRB 170114A (*in units of erg/cm$^2$ in 10-1000 keV).}\label{tab:res_summary}
\begin{tabular}{|c|c|c|c|c|c|c|c|}
  \hline
  GRB & T90 (s) & Fluence* & PD & Prob(PD$<$2\%) & PD$_{\rm up}$(99\%) & PA(deg.) & PA Change \\ \hline\hline
  161218A& 6.76 & $1.25\times10^{-5}$ & 9\% & 9\% & 45\% & 40 & No \\ \hline
  170101A& 2.82 & $1.27\times10^{-5}$ & 8\% & 13\% & 31\% & 164 & No \\ \hline
  170127C& 0.21 & $7.4\times10^{-6}$ & 11\% & 5.8\% & 67\% & 38 & Unknown \\ \hline
  170206A& 1.2 & $1.34\times10^{-5}$ & 10\% & 12\% & 31\% & 106 & No  \\ \hline
  170114A& 8.0 & $1.93\times10^{-5}$ & 4\% & 14\% & 28\% & 164 & Yes  \\ \hline
  170114Ap1& N$/$A & N$/$A & 15\% & 8\% & 43\% & 122 & N$/$A  \\ \hline
  170114Ap2& N$/$A & N$/$A & 41\% & 0.49\% & 74\% & 17 & N$/$A  \\ \hline
\end{tabular}
\end{table}

\clearpage

\section*{Methods}

\subsection*{Sample selection}

The GRBs analyzed in this paper are selected from the POLAR GRB catalog, presented in ref. ~\cite{Xiong2017}, based on the following criteria:

\begin{itemize}
 \item The GRB has been observed by detectors other than POLAR and good measurement of both the spectrum and location are provided by other instruments.
 \item The fluence of the GRB, as provided by other instruments in the $10-1000\,\mathrm{keV}$ energy range, exceeds $5\times10^{-6}\mathrm{erg/cm^2}$.
 \item The incoming angle with respect to the POLAR instrument zenith, $\theta$, is below $45^\circ$.
\end{itemize}

The first selection criterion minimizes the systematic errors in the polarization results induced by uncertainties in the spectrum and location of the GRB. The second and third criteria ensure a large number of events and a large modulation factor ($M_{100}$, the amplitude of the modulation measured for a $100\%$ polarized flux) and therefore a high statistical significance of the measurement. The third criterion furthermore reduces the influence of photons scattered off the objects in the vicinity of POLAR, thereby removing additional systematic errors.

This selection resulted in the following six GRBs eligible for study: 161129A, 161218A, 170101A, 170114A, 170127C and 170206A. The light curves of these GRBs can be found in ref. ~\cite{light_curves}. GRB 161129A was however preceded by a large solar flare. It is therefore not selected as systematics induced by potential low level irradiation from the last part of the flare require further studies. Therefore, only the latter five GRBs are included for this work. The light curves of the five GRBs are shown in Supplementary Figure 1. In these light curves only events are selected where at least two bars are above trigger level as only these events can be used for polarization studies. Details of the event selection including the trigger algorithm can be found in ref. ~\cite{LI2018}. Furthermore the different time bins studied for GRB 170114A, discussed later in this document, are indicated in the bottom right panel of the figure.

\subsection*{Methodology of GRB polarization analysis}

A full detailed instrument calibration of the POLAR detector in-orbit is a prerequisite for the polarization study. The results of such a study are presented in ref.~\cite{LI2018}. The polarization analysis is performed by producing one modulation curve from the measured data. This modulation curve is fitted to a simulated instrument response consisting of 6060 simulated modulation curves to find the best fitting polarization degree (PD) and polarization angle (PA), described in more detail in the next section.

The simulated response is produced using the Monte Carlo (MC) software as presented in ref.~\cite{Kole2017}. It should be noted that some updates to the MC software have been implemented since the publication of ref.~\cite{Kole2017} based on a deeper analysis of the in-orbit data. These updates mainly concern the inclusion of a detailed mass model of the Tiangong-2 (TG-2) space lab, on which POLAR is mounted, and a better parameterization of the non-linearity of the energy gain based on the findings presented in ref.~\cite{LI2018}. The measured and simulated data are processed using the methods described in ref.~\cite{LI2018}.

The event selection as well as the data processing of both the measured and simulated data is equivalent. Events selected for analysis consist of all the clean events as described in ref.~\cite{LI2018}. Additionally no channels with an energy deposition above overflow are selected. Finally the two channels in POLAR with the highest energy depositions, as measured in keV, are selected. The angle between the two bars is taken to be the scattering angle. The first bar of the two is that with the highest energy deposition within the whole event. The second bar is selected as that with the highest energy deposition among those bars not adjacent to the first bar. If the second channel is not found with the non-adjacent criterion the event is not selected. The non-adjacent criterion is used to improve $M_{100}$ and to reduce the influence of the fluctuations of crosstalk signals. In order to reduce the effect of the energy threshold non-uniformity, an additional energy threshold of 15 keV is applied in the event selection on the energy deposition of the second channel.

For the measured modulation curve the selected signal time intervals are based on the T90 values measured by POLAR. The background time interval is selected as two time intervals which are respectively before and after the GRB as described in more detail in the supplementary material and indicated in Supplementary Figure 1. The modulation curve of the background is subtracted from the modulation curve produced using data in the selected signal time interval. Here the relative difference between the length of the signal and background intervals is taken into account, as well as the error propagation for each bin. As the background is relatively stable, as illustrated in Supplementary Figure 6, the influence of the background selection on the final results was found to be negligible in a dedicated study described in detail in a dedicated section in the supplementary materials. The results are summarized in Supplementary Tables 2 and 3.

For each GRB a range of simulations is performed. In each the GRB is simulated using photons emitted from a circular plane with a radius of 250 mm, with the incident direction calculated based on the best available measured location for the GRB from other instruments. The influence of systematic uncertainties in the location of the GRB on our final results was found to be small through a dedicated study described in a dedicated section of the Supplementary Materials. Its results are summarized in Supplementary Table 5. The size of the emitting plane is sufficient to illuminate the full instrument as well as the materials surrounding POLAR. It should furthermore be noted that it was found that, by once removing them in the simulations, the materials surrounding POLAR have no significant influence on the final polarization measurement for the studied GRBs. 

The energy spectrum of the simulated photons follows a Band function~\cite{Band1993} using the best published parameters for the GRB. The influence of the choice of the used spectrum, when different spectra were available, was found to be negligible through a study described in a dedicated section of the Supplementary Materials. The results of this study are summarized in Supplementary Table 4. It should be noted here that the spectrum of GRB 170127C is best fitted using a single cutoff power-law and is therefore also modeled as such in the POLAR analysis. The lower limit for the energy range for simulated photons is 10 keV, below which the effective area of POLAR becomes negligible. The upper limit is set to 1000 keV for all GRBs except for 170127C, where this is extended up to 1500 keV where, according to the spectra provided by Fermi-GBM, the flux drops below $1\%$ of that at 10 keV. The photons are then simulated with an origin from a random position in the circular plane and a momentum vector based on the GRB location with respect to POLAR. For each GRB a total of 61 simulations are performed each with 5 million photons. A total of 60 simulations are performed for the 100\% polarized photons, differing in the polarization angles (PA) with step sizes of 3 degrees. One simulation is performed for the unpolarized photons. The modulation curves of the polarization degrees (PD) between 0--100\% can be generated by mixing those of the unpolarized flux and the 100\% polarized flux. Therefore, using the modulation curves of the 61 simulations, a total of 6060 different simulated modulation curves was generated in the 2-D plane of PA and PD with a step of 1\% in the PD direction and 3 degrees in the PA direction. The modulation curves produced this way, as well as those from the measured data, are normalized by the total number of events within the modulation curve. Subsequently the $60 \times 101$ array of simulated modulation curves are used to find the best fitted PA and PD for the measured modulation curve using the least $\chi^2$ method as discussed in the next section. This array of simulated modulation curves is also used for the Bayesian analysis method, discussed later in this document, to generate the posterior probability distribution of the true PA and PD. 

The results of this analysis when applied on full GRBs is detailed in Supplementary Table 7 while the input parameters and their origins are summarized in Supplementary Table 6. 

\subsection*{Least $\chi^2$ analysis method}

In order to determine the PA and PD of each GRB the measured modulation curve is fitted with the simulated instrument response using the least $\chi^2$ method. One $\chi^2$ value between the measured modulation curve and one of the 6060 simulated modulation curves as mentioned above can be calculated using Eq~\eqref{equ:chi2},
\begin{equation}\label{equ:chi2}
\chi^2=\sum_{i=1}^n \frac{(X_i-Y_i)^2}{\varepsilon_i^2+\sigma_i^2},
\end{equation}
where $X_i$ and $Y_i$ are the counts of the bins of the measured and simulated modulation curves respectively, $\varepsilon_i^2$ and $\sigma_i^2$ are the uncertainties of $X_i$ and $Y_i$ respectively, and $n$ is the number of bins in the modulation curves. Note that both the measured and simulated modulation curves are normalized before calculating the $\chi^2$ value. The uncertainties of the bins in the measured modulation curve are taken to be the statistical errors. On top of the statistical uncertainties in the simulated modulation curves these also contain systematic uncertainties resulting both from the spectral and location parameters of the GRB as well as from uncertainties in the calibration parameters used in the simulations. The details of the systematic error determination is described in the supplementary material. 

Using the 6060 simulated modulation curves and the single measured modulation curve with their corresponding errors the $60 \times 101$ array of $\chi^2$ values corresponding to different PAs and PDs can be calculated. As an example Supplementary Figure 2 shows the map of $\Delta\chi^2 = \chi^2 - \chi^2_{\rm min}$ for GRB 170206A. The best fitted PA and PD of this GRB are those corresponding to the $\chi^2_{\rm min}$.

As discussed in ref.~\cite{Avni1976}, the confidence area of the PA--PD measurement can be determined using the value of $\Delta\chi^2$. In Supplementary Figure 2, the three black contours from left to right correspond to $\Delta\chi^2 = 2.28,4.61$ and $9.21$ which are the upper quantiles with probabilities 32\%, 10\% and 1\% for the $\chi^2$ distribution of 2 degree of freedom (d.o.f.). In order to get the upper limit of PD without considering the value of PA, the upper quantile of the 1 d.o.f. $\chi^2$ distribution is used. The red contour corresponds to $\Delta\chi^2 = 6.64$ which is the upper quantile with probability 1\% for the 1 d.o.f. $\chi^2$ distribution. The maximum PD on the red contour is the upper limit of PD with confidence level 99\%, which is approximately 34\%.

\subsection*{Bayesian analysis method}

Suppose $\bm{A}=(\hat{p},\hat{\phi})$ is the measurement of PD and PA using the least $\chi^2$ method as discussed in the previous section, and $\bm{B}_i=(p,\phi)$ is the true PD and PA of the source. Then $P(\bm{A}|\bm{B}_i)$ is the probability that the measurement of PD and PA is $\bm{A}$ under the condition that the true PD and PA of the source is $\bm{B}_i$. If the prior probability of the true PD and PA of the source is $P(\bm{B}_i)$, the posterior probability of $\bm{B}_i$, under the condition that the measurement of PD and PA is $\bm{A}$, can be acquired using the Bayesian equation as presented by Eq.~\eqref{equ:bayes_lisan}.
\begin{equation}\label{equ:bayes_lisan}
P(\bm{B}_i|\bm{A})=\frac{P(\bm{A}|\bm{B}_i)P(\bm{B}_i)}{\sum_{j=1}^n P(\bm{A}|\bm{B}_j)P(\bm{B}_j)}.
\end{equation}
Here $P(\bm{A}|\bm{B}_i)$ for different bins of $\bm{B}_i$ can be acquired by performing a series of measurement simulations, the procedure for which is described in the supplementary material. Some examples of the distribution of $P(\bm{A}|\bm{B}_i)$ for GRB 170206A are shown in the Supplementary Figure 3. As there is no information of the true PD and PA of the source before the measurement, the prior probability of $\bm{B}_i$ is taken to be uniform in the range of 0\% and 100\% for PD and $0^\circ$ and $180^\circ$ for PA and 0 outside this range. After producing the distribution of $P(\bm{A}|\bm{B}_i)$ for all different bins of $\bm{B}_i$, using simulations, the posterior probability distribution of the true PA and PD of the source under the condition of the single measurement can be directly calculated with Eq.~\eqref{equ:bayes_lisan}. The Supplementary Figure 4 shows the result of $P(\bm{B}_i|\bm{A})$ for GRB 170206A, and Figure \ref{fig:comb_post} show the results of all the 5 selected GRBs.

After integrating the 2-D posterior probability distribution of PD and PA, as shown in Figure \ref{fig:comb_post} along the PA direction, a 1-D posterior probability distribution of PD is acquired as illustrated in Supplementary Figure 5. This 1-D distribution can be used to find the upper limit of PD with a certain credible level by integrating it from right to left and to find the lower limit by integrating it from left to right, as shown in Figure \ref{fig:comb_CDF}. From Figure \ref{fig:comb_CDF}(b), it can be seen that the upper limit of PD with credible level 99\% is about 31\%, which is very similar to that obtained with the least $\chi^2$ method.

\subsection*{Evolution of polarization angle}

For all GRBs with exception of 170127C, which is too short and lacks statistics, a time binned analysis is also performed. For this purpose the data from the GRBs is split into equal time intervals. The GRBs 161218A, 170101A and 170206A are split into two intervals. GRB 170114A, which consists of a bright peak lasting approximately 6 seconds followed by 3 seconds with few events, is divided into three time bins. These time intervals for 170114A are indicated in the bottom right panel of Supplementary Figure 1. For this GRB only the  two time bins in the peak are discussed here as the MDP for the third bin is significantly larger than 100\%, indicating that no constraining measurement on the polarization properties of this interval are possible.

The analysis is performed using two different approaches for each GRB. First the data from both time bins are analyzed independently following the same procedure as that applied in the analysis of the time integrated GRBs. For the two time bins the incoming angles as used in the simulations is corrected for the minor changes induced by the rotation of POLAR with respect to the GRB during the burst. The second applied method is the same as that used by the GAP collaboration for GRB 100826 ~\cite{GAP}. In this analysis the two time intervals are analyzed simultaneously while the PD of both intervals is forced to be equal, while the PA are allowed to vary. This study therefore takes the assumption that the PD is constant throughout the GRB while the PA can vary with time.

For GRBs 161218A, 170101A and 170206A the analyses using both methods is found to give consistent results for both time intervals to the analysis of the full time integrated GRB. No significant change in the angle is therefore found during these GRBs. It should however be noted that this analysis is only performed when dividing the full GRB in two equal time intervals; we can not exclude that the polarization angle varies on shorter time scales, since the data of POLAR lack statistics to do such studies.

For GRB 170114A a significant evolution of the polarization angle is found using both analysis methods. In the analysis the changes in the incoming angle of the GRB during the burst as well as spectral evolution of the GRB is taken into account. Such spectral evolution was not reported in the GCN by Fermi-GBM~\cite{GCN_170114A}, the only other instrument apart from POLAR that reported a detection of this GRB. However, analysis of the Fermi-GBM data shows that a significant spectral evolution occurred during the GRB. Using independent analysis of both bins and using new spectra for the two time bins resulted in the modulation curves presented in Supplementary Figure 7 and 9 and the $\Delta\chi^2$ distributions shown in Supplementary Figures 8 and 10. The first time interval is consistent with an unpolarized or lowly polarized flux and gives an upper limit of a polarization degree of 42\%, while the second interval only has a $0.49\%$ probability for PD lower than $2\%$. The 99\% upper limit for the polarization degree of the second part is 74\%.

It was furthermore found that the PD of both time intervals is compatible within $1\sigma$ with a PD around 25\%, albeit with very different polarization angles. We can therefore perform the second analysis approach where PD for both bins is forced to be equal. The result of this study on GRB 170114A is a PD of 24\% throughout the GRB while the PA of the first time bin is $116^\circ$ and that of the second time bin is $11^\circ$ with a $\chi^2/$NDF of $59.6/55$. The resulting $\Delta\chi^2$ map for the first time bin is shown in Supplementary Figure 11. The $\Delta\chi^2$ map of the second time bin is the same as this but shifted along the PA axis. The results exclude an unpolarized flux with a 99.2\% confidence level.

The statistics for this GRB allow the peak to be divided into three time bins, each of 2 seconds. Using the 3 time bins the same analysis results in a PD of 28\% with PA=$98^\circ$ for the first time bin, PA=$152^\circ$ for the second time bin and PA=$38^\circ$ for the final time bin. The $\chi^2/$NDF for this analysis is 27.6/29, note that the number of bins in the modulation curves is reduced here in order to allow for sufficient statistics in each bin. The final result in the form of a $\Delta\chi^2$ map for the first time bin is shown in Supplementary Figure 12. This analysis results in a higher exclusion level with PD=28\% with a $1\sigma$ error of 9\% and excludes an unpolarized flux with 99.7\% confidence.

\section*{Data Availability Statement}
All the data that support the plots within this paper and other findings of this study are available from the POLAR Collaboration (merlin.kole@unige.ch) upon reasonable request.

\renewcommand{\refname}{References}

\newpage

\section*{Supplementary Information}

\section{Compilation of all previously reported GRB polarization results}

In Supplementary Table~\ref{tab:all_GRBs} the POLAR measurement results are placed together with all previously published GRB measurements by other instruments which can also be found in ref. ~\cite{Covino:2016cuw}.

\begin {table}[!ht]
\begin{center}
\caption{\textbf{Summary of all published GRB prompt emission polarization measurements.} The quoted errors or limits are at the $1\sigma$ level unless otherwise specified. (*The result of GRB 170114A is from the time binned analysis with 3 time bins where PD was assumed to be constant during the burst.)}
  \begin{tabular}{ | l | c | c | c |}
    \hline
    \textbf{GRB} & \textbf{Instr./Sat.} & \textbf{Pol. ($\%$)} & \textbf{Ref.} \\ \hline\hline
    170206A & POLAR & $<31\%\,$($99\%$ confidence) & This Paper  \\ \hline
    170127C & POLAR & $<68\%\,$($99\%$ confidence) & This Paper  \\ \hline
    170114A & POLAR & $<28\%\,$($99\%$ confidence) & This Paper  \\ \hline
    170114A & POLAR & $ 28\pm 9\%$ & This Paper, time binned*  \\ \hline
    170101A & POLAR & $<30\%\,$($99\%$ confidence) & This Paper  \\ \hline
    161218A & POLAR & $<41\%\,$($99\%$ confidence) & This Paper  \\ \hline\hline
    160530A & COSI & $<46\%\,$($90\%$ confidence) &ref. ~\cite{COSI2}  \\ \hline\hline
    110721A & GAP/IKAROS & $84^{+16}_{-28}$ &ref. ~\cite{GAPGRB2} \\ \hline
    110301A & GAP/IKAROS & $70\pm22$ &ref. ~\cite{GAPGRB2}  \\ \hline
    100826A & GAP/IKAROS & $27\pm11$ &ref. ~\cite{GAP} \\ \hline \hline
    021206 & RHESSI & $80\pm20$ &ref. ~\cite{RHESSI2}  \\ \hline
    021206 & RHESSI & $41^{+57}_{-44}$ &ref. ~\cite{RHESSI1}  \\ \hline \hline

    140206A & IBIS/INTEGRAL & $\ge28$ ($90\%$ confidence)&ref. ~\cite{IBIS1}  \\ \hline
    061122 & IBIS/INTEGRAL & $\ge33$ ($90\%$ confidence)&ref. ~\cite{IBIS3} \\ \hline
    041219A & IBIS/INTEGRAL & $\le4 /  43\pm25$ &ref. ~\cite{IBIS2}  \\ \hline
    041219A & SPI/INTEGRAL & $98\pm33$ &ref. ~\cite{SPI}  \\ \hline \hline
    960924 & BATSE/CGRO & $\ge50$ &ref. ~\cite{BATSE1}  \\ \hline
    930131 & BATSE/CGRO & $\ge35$ &ref. ~\cite{BATSE1}  \\ \hline

  \end{tabular}
\label{tab:all_GRBs}
\end{center}
\end {table}

\newpage

\section{Sample selection supplements}

\begin{figure}[!ht]
\centering
\includegraphics[height=12.0cm]{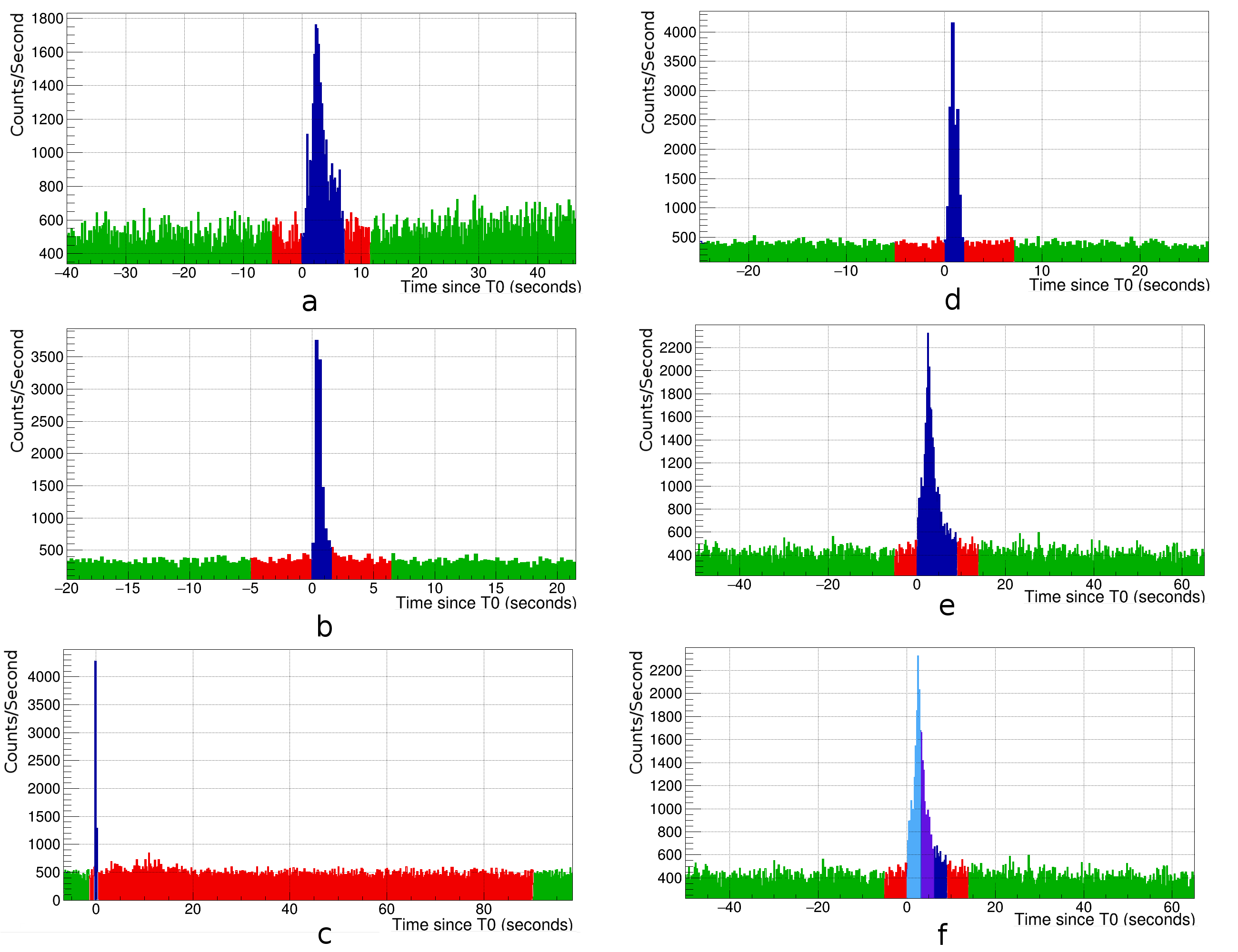}
\caption{\textbf{Light curves of all the studied GRBs.} The selected signal region is shown in blue, the background region in green and the non-used data in red. Left from top to bottom are 161218A (a), 170101A (b) and 170127C (c); Right from top to bottom are 170206A (d), 170114A (e) and 170114A with additional details where the different time bins of 170114A are indicated with different shades of blue (f).}\label{fig:lc}
\end{figure}

\clearpage
\section{Systematic Error Studies}

The simulated modulation curves contain systematic errors, which are dominant for these curves due to very small statistical error resulting from the large statistics used for each simulation. In order to understand the scale of the systematic errors on the simulated modulation curves, a set of 1000 simulations is performed for one of the studied GRBs: GRB 170206A. The systematic error is expected to come from two main sources. Firstly the simulated modulation curves are based on the incoming spectrum and direction of the GRB. The parameters for the spectra and incoming direction of the photons are acquired from measurements from other instruments, such as Fermi-GBM, Konus-Wind and Swift-BAT, and have significant errors which should be taken into account. The second source of systematic errors stems from the errors in the calibration parameters used by the simulations. As described in ref.~\cite{LI2018}, the calibration parameters whose uncertainties dominate the effect on the modulation curve are the gain parameters.

In each of the 1000 simulations in this systematics study the input parameters of the energy gain of each channel, the spectral parameters and the incident angle of the GRB are randomized within their errors. This results in 1000 modulation curves each with a high statistical significance. The uncertainties on the different input parameters are found to result in an average relative variation of 4\% on the bin content after statistical variations are subtracted. This procedure is repeated for three different configurations of PD and PA, and no significant difference is found on this variation of bin counts induced by the errors of input parameters. The systematic uncertainties are therefore taken into account by adding a 4\% relative error to each bin in the simulated modulation curves. This systematic error being relatively small is partly a result of the selection procedure of the GRBs in this analysis. The selection criteria limit the systematic errors coming from the spectral and location uncertainties. The systematics are further reduced as a result of the careful calibration of the POLAR instrument. Although the systematic errors are not negligible in the analysis, the statistical errors in the measured modulation curve dominate for all the GRBs studied here.

\newpage
\section{Least $\chi^2$ analysis method supplements}

\begin{figure}[!ht]
  \centering
    \includegraphics[width=12 cm]{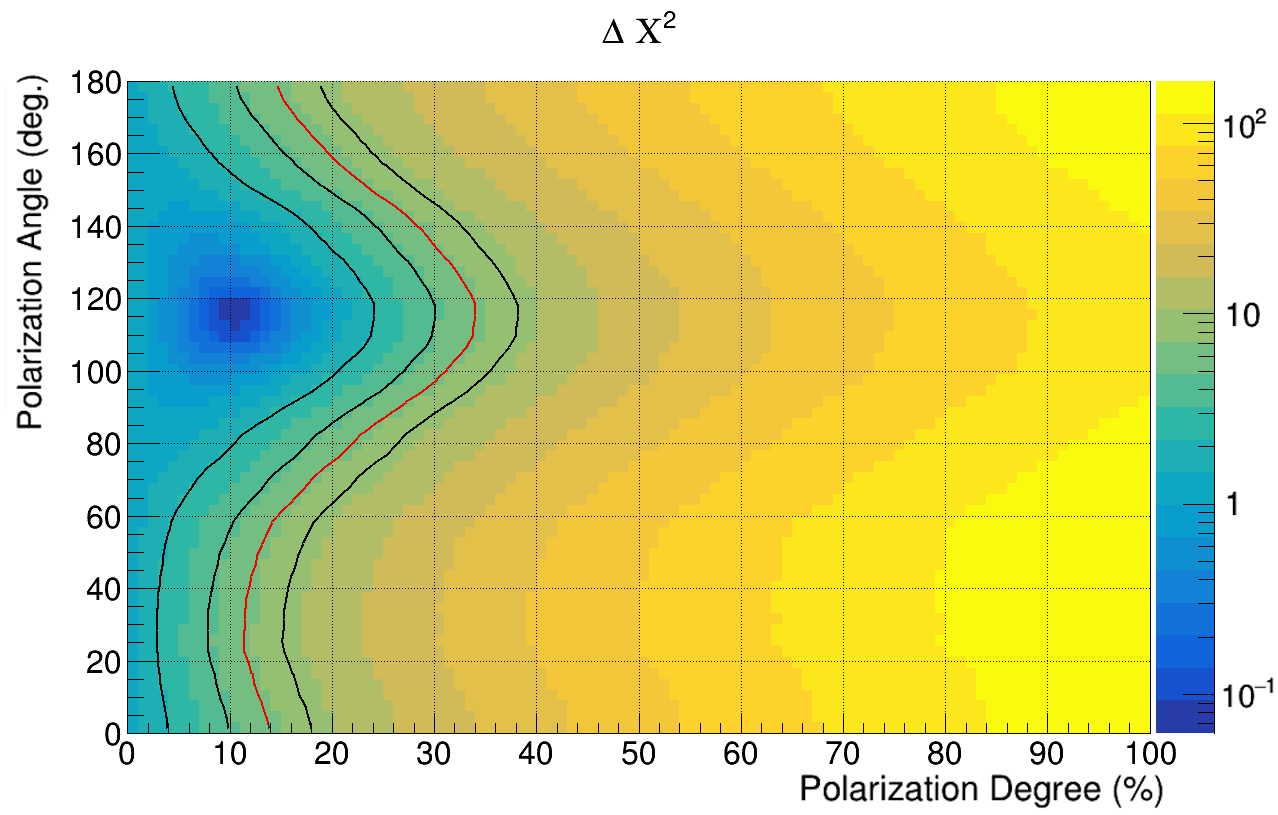}
  \caption[$\Delta\chi^2$ map for GRB 170206A]
 {\textbf{$\Delta\chi^2$ map for GRB 170206A.} The $\Delta\chi^2$ values of the fits are shown as a function of PD and PA for GRB 170206A. Confidence areas of PA and PD are additionally drawn. The three black contours from inner to outer show the PA--PD confidence area with confidence levels 68\%, 90\% and 99\%. The red contour with $\Delta\chi^2=6.635$ is used to determine the upper limit of PD with confidence level 99\%.
}
\label{fig:DChi2_170206A}
\end{figure}

\newpage
\section{Bayesian analysis method supplements}\label{sec:bayes}

Suppose $\bm{A}=(\hat{p},\hat{\phi})$ is the measurement of PD and PA using the least $\chi^2$ method as discussed in the methods section, and $\bm{B}_i=(p,\phi)$ is the true PD and PA of the source. Then $P(\bm{A}|\bm{B}_i)$ is the probability that the measurement of PD and PA is $\bm{A}$ under the condition that the true PD and PA of the source is $\bm{B}_i$. If the prior probability of the true PD and PA of the source is $P(\bm{B}_i)$, the posterior probability of $\bm{B}_i$ under the condition that the measurement of PD and PA is $\bm{A}$ can be acquired using the Bayesian equation as presented by Eq.~\eqref{equ:bayes_lisan},
\begin{equation}\label{equ:bayes_lisan}
P(\bm{B}_i|\bm{A})=\frac{P(\bm{A}|\bm{B}_i)P(\bm{B}_i)}{\sum_{j=1}^n P(\bm{A}|\bm{B}_j)P(\bm{B}_j)}.
\end{equation}
It should be noted here that Eq.~\eqref{equ:bayes_lisan} is the Bayesian equation of discrete type. However the values of PD and PA are continuous. The Bayesian equation of the continuous type can be presented by Eq.~\eqref{equ:bayes_lianxu},
\begin{equation}\label{equ:bayes_lianxu}
h(\bm{B}|\bm{A})=\frac{f(\bm{A}|\bm{B})\pi(\bm{B})}{\int_{\bm{B}}f(\bm{A}|\bm{B})\pi(\bm{B}) d\bm{B}}.
\end{equation}
However, in numerical analysis the continuous values are always discretized. Therefore, Eq.~\eqref{equ:bayes_lianxu} can be converted to Eq.~\eqref{equ:bayes_lisan} by Eq.~\eqref{equ:mianyuan},
\begin{equation}\label{equ:mianyuan}
f(\bm{A}|\bm{B})d\bm{A} \Rightarrow P(\bm{A}|\bm{B}_i) \quad \pi(\bm{B})d\bm{B} \Rightarrow P(\bm{B}_i) \quad h(\bm{B}|\bm{A})d\bm{A} \Rightarrow P(\bm{B}_i|\bm{A}),
\end{equation}
where $d\bm{A}$ and $d\bm{B}$ are the small but limited bins in the 2-D plane of PD and PA. In Eq.~\eqref{equ:mianyuan} the probability density function like $f(\bm{A}|\bm{B})$ is converted to the probability that the measurement value or the true value occurs within one specific PA--PD bin by multiplying the bin size. Therefore Eq.~\eqref{equ:bayes_lisan} can be directly used in the Bayesian analysis after binning the PA and PD.

\begin{figure}[!ht]
\centering
\subfigure[]{\includegraphics[height=4.5cm]{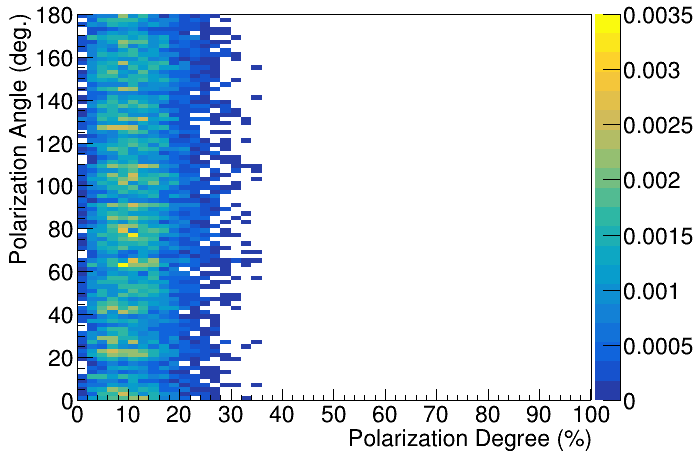}}
\hspace{2mm}
\subfigure[]{\includegraphics[height=4.5cm]{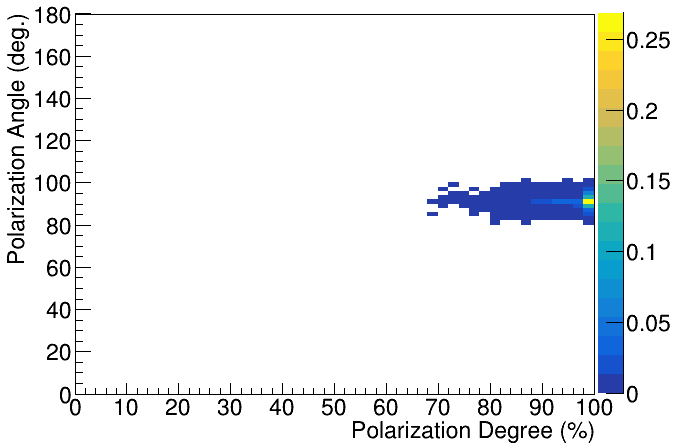}}
\subfigure[]{\includegraphics[height=4.5cm]{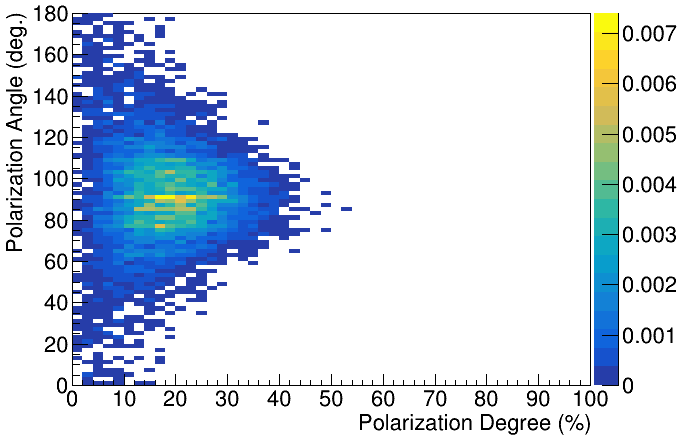}}
\hspace{2mm}
\subfigure[]{\includegraphics[height=4.5cm]{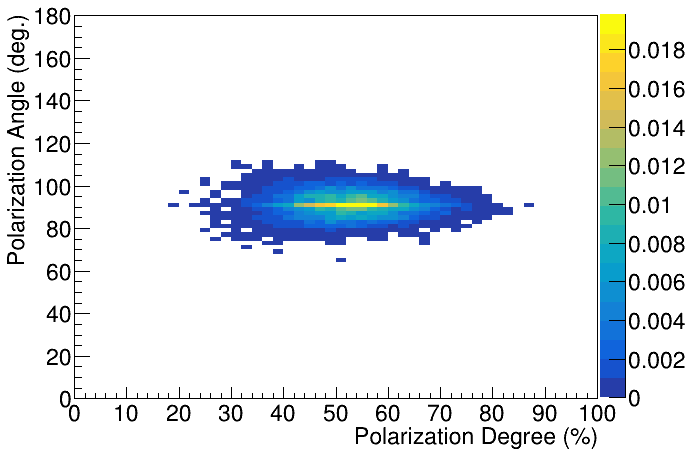}}
\subfigure[]{\includegraphics[height=4.5cm]{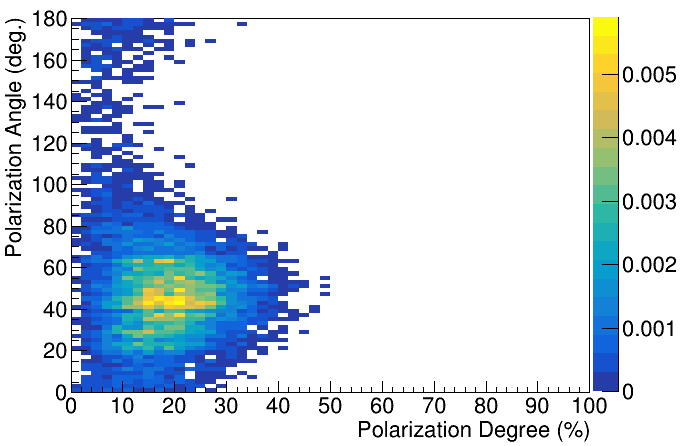}}
\hspace{2mm}
\subfigure[]{\includegraphics[height=4.5cm]{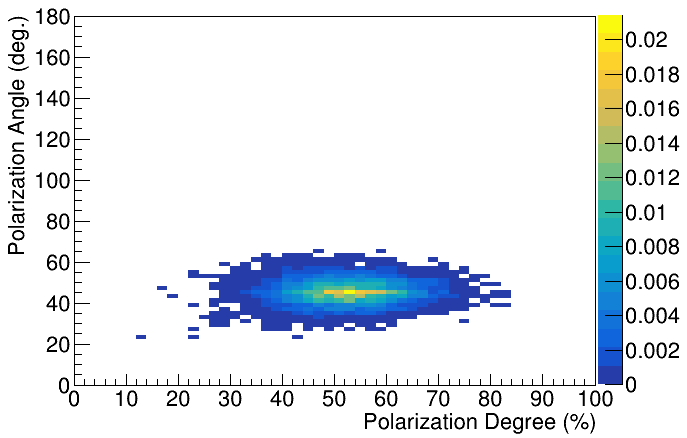}}
\caption{\textbf{Illustrative probability distributions of the measurement of PA and PD under the condition of different true PA and PD.} These distributions are produced using the statistical quantities of the modulation curve of GRB 170206A. The true PA and PD in different panels are respectively (a) PA = 91$^\circ$, PD = 1\%; (b) PA = 91$^\circ$, PD = 99\%; (c) PA = 91$^\circ$, PD = 15\%; (d) PA = 91$^\circ$, PD = 51\%; (e) PA = 45$^\circ$, PD = 15\%; (f) PA = 45$^\circ$, PD = 51\%.}\label{fig:pa_pd_prob}
\end{figure}

In order to get $P(\bm{A}|\bm{B}_i)$ for different $\bm{B}_i$, a series of simulations is required. The simulation here is called the measurement simulation, where the modulation curve is sampled with the same statistical quantity as that of the real data. The modulation curve is sampled from a high statistics modulation curve. The addition and subtraction of the background is also taken into account. For this purpose the modulation curve of the background is sampled using the same statistical quantity as that of the background within the selected GRB signal time interval and added to the sampled modulation curve of the signal. Subsequently the sampled modulation curve produced this way is analyzed using the same method as that used by the modulation curve of the real data. This includes background subtraction and finding the best fitted PA and PD using the least $\chi^2$ method. For each $\bm{B}_i$ 10,000 such measurement simulations are performed to get the distribution of $P(\bm{A}|\bm{B}_i)$. Here in the PA direction 90 bins are used while in the PD direction 50 bins are used, therefore totally $90 \times 50 \times 10000 = 45,000,000$ measurement simulations are performed for each GRB. Supplementary Figure~\ref{fig:pa_pd_prob} shows some examples of the distribution of $P(\bm{A}|\bm{B}_i)$ after normalization. These are produced using measurement simulations with different values of $\bm{B}_i$ for the case of GRB 170206A.

There is only one real measurement of PD and PA which is $\bm{A}$. As there is no information of the true PD and PA of the source before the measurement, the prior probability of $\bm{B}_i$ can be assumed as uniform in the range of 0\% and 100\% for PD and $0^\circ$ and $180^\circ$ for PA and 0 outside this range. Therefore, after the distributions of $P(\bm{A}|\bm{B}_i)$ for all different bins of $\bm{B}_i$ are produced by the measurement simulation, the posterior probability distribution of the true PA and PD of the source under the condition of the single measurement can be directly calculated with Eq.~\eqref{equ:bayes_lisan}, which is shown in Supplementary Figure~\ref{fig:post_170206A} for GRB 170206A.

\begin{figure}[!ht]
\centering
\includegraphics[height=7.0cm]{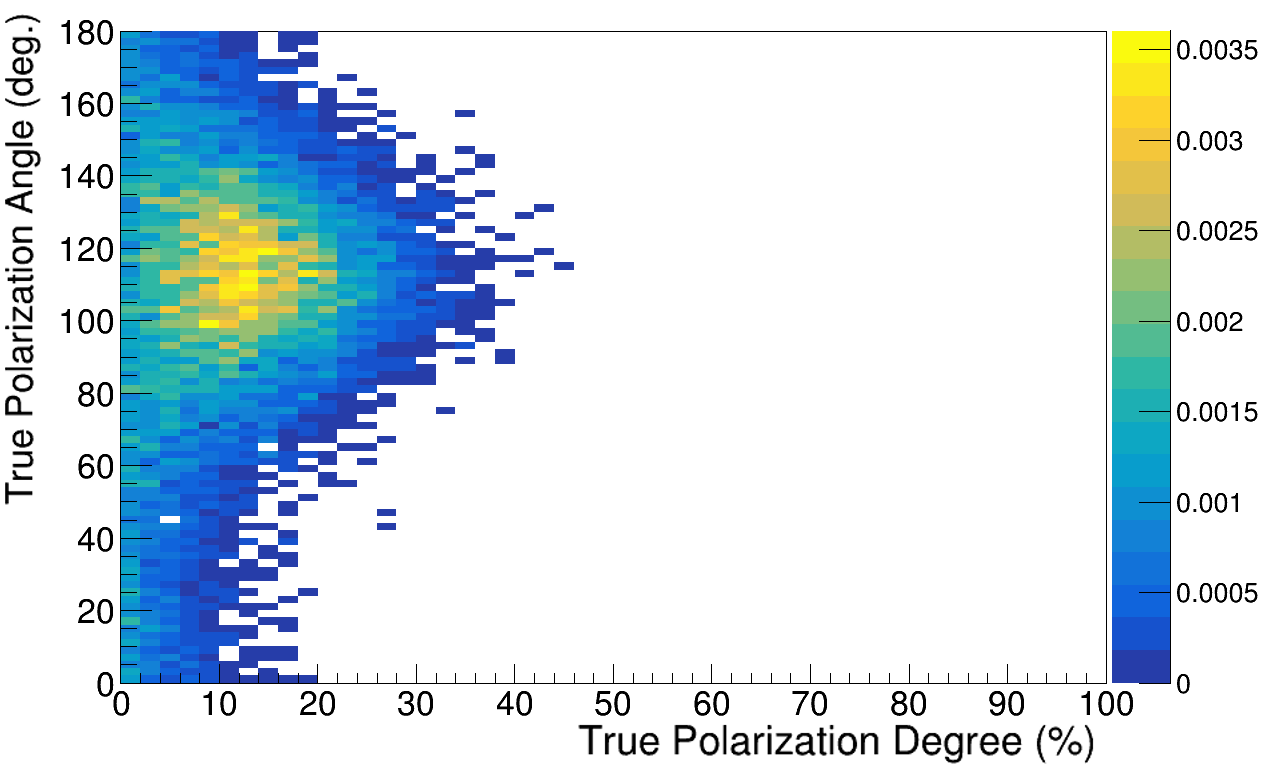}
\caption{\textbf{Posterior probability distribution of the true PA and PD for GRB 170206A.}}\label{fig:post_170206A}
\end{figure}

After integrating the 2-D posterior probability distribution of PD and PA as shown in Supplementary Figure~\ref{fig:post_170206A} along the PA direction, a 1-D posterior probability distribution of PD is acquired. This can be used to find the upper limit of PD with a certain credible level as shown in Supplementary Figure~\ref{fig:pd_post}. The vertical orange line in Supplementary Figure~\ref{fig:pd_post} shows that the upper limit of PD with credible level 99\% is at approximately 31\%, which is very similar to that obtained with the least $\chi^2$ method as discussed in the methods section of the main paper. The cumulative probability, such as that shown in the paper, can be calculated by integrating the posterior probability distribution over PD. Supplementary Figure~\ref{fig:pd_post}(b) shows the cumulative probability integrating the posterior probability distribution as shown in Supplementary Figure~\ref{fig:pd_post}(a) over PD from right to left and Supplementary Figure~\ref{fig:pd_post}(b) presents the upper limit credible level of PD. The lower limit credible level of PD can be presented by the cumulative probability integrating the posterior probability distribution of PD from left to right as shown in the paper.

\begin{figure}[!ht]
\centering
\subfigure[]{\includegraphics[height=4.7cm]{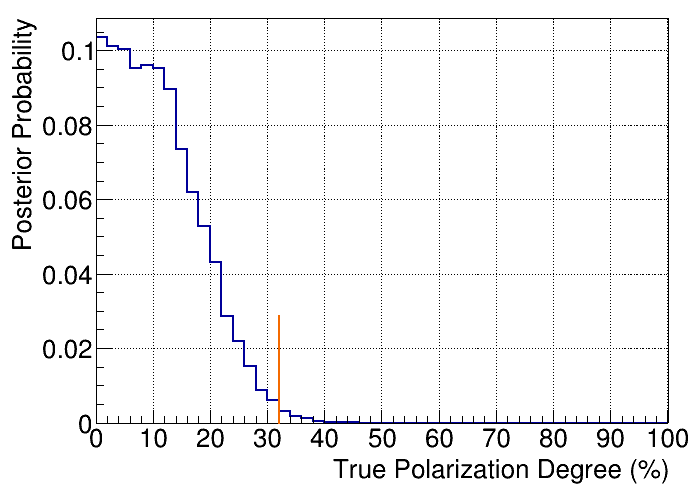}}
\subfigure[]{\includegraphics[height=4.7cm]{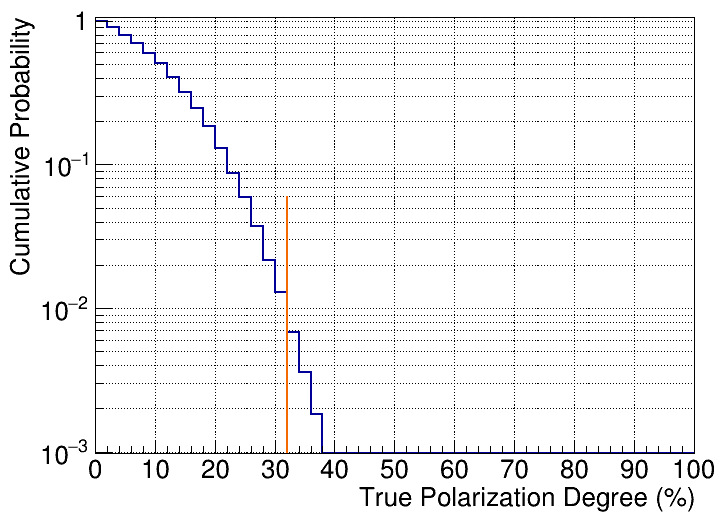}}
\caption{\textbf{Illustrative posterior and corresponding cumulative probability distribution.} The blue histograms show the posterior probability distribution of the true PD for GRB 170206A and the corresponding cumulative probability distribution integrating from right to left. The vertical orange line shows the upper limit of PD with credible level 99\%, which is approximately 31\%.}\label{fig:pd_post}
\end{figure}

\clearpage
\section{Input parameter influence studies}
\subsection{Background selection influence}\label{sec:bg}

In order to test the dependency of the final polarization result on the selected background time interval, the stability of the background modulation is tested. The background time interval selected for analysis is in total at least 10 times longer than the signal time interval. Half of the interval is taken before the GRB and ends 5 seconds before the start of the GRB. The second half starts 5 seconds after the end of the selected signal time interval. It should be noted here that GRB 170127C comprised one bright burst of 0.2 seconds followed by a weak  emission of approximately 90 seconds according to Konus-Wind observations~\cite{KONUS_GCN}. For the analysis here only the bright peak is analyzed and the gap interval after the GRB is therefore set to 90 seconds. All the selected background time intervals can be seen in Supplementary Figure~\ref{fig:lc}.

An example of the (non-normalized) modulation curves of the background time intervals after GRB 170114A with a gap interval of 5 seconds is shown in Supplementary Figure~\ref{bg_2} together with the background modulation curve for a gap interval 90 seconds. The figure indicates that the background modulation is stable at least within a reasonable time around the GRB. In order to further study the effect of selecting different gap intervals (the time interval indicated in red in Supplementary Figure~\ref{fig:lc}), the analysis of GRB 170114A is performed three times with different lengths for the gap interval. The results are summarized in Supplementary Table~\ref{tab:gap_bg} with the best fitting PD, PA and the corresponding value of the $\chi^2$/NDF.

\begin {table}[!ht]
\small
\begin {center}
\caption{\textbf{The resulting PD and PA values together with the corresponding $\chi^2$/NDF acquired when selecting different intervals between the signal and background time intervals.}}
\begin{tabular}{| l | c | c | c |}
  \hline			
  time interval & PD. (\%) & PA (deg.) & minimum $\chi^2$/NDF \\
  \hline			
  5 & 4\% & $164^\circ$ & 1.19 \\
  45 & 3\% & $164^\circ$ & 1.08 \\
  90 & 3\% & $164^\circ$ & 1.18 \\
\hline
\end{tabular}
\label{tab:gap_bg}
\end{center}
\end{table}

\begin{figure}[!ht]
  \centering
    \includegraphics[width=12 cm]{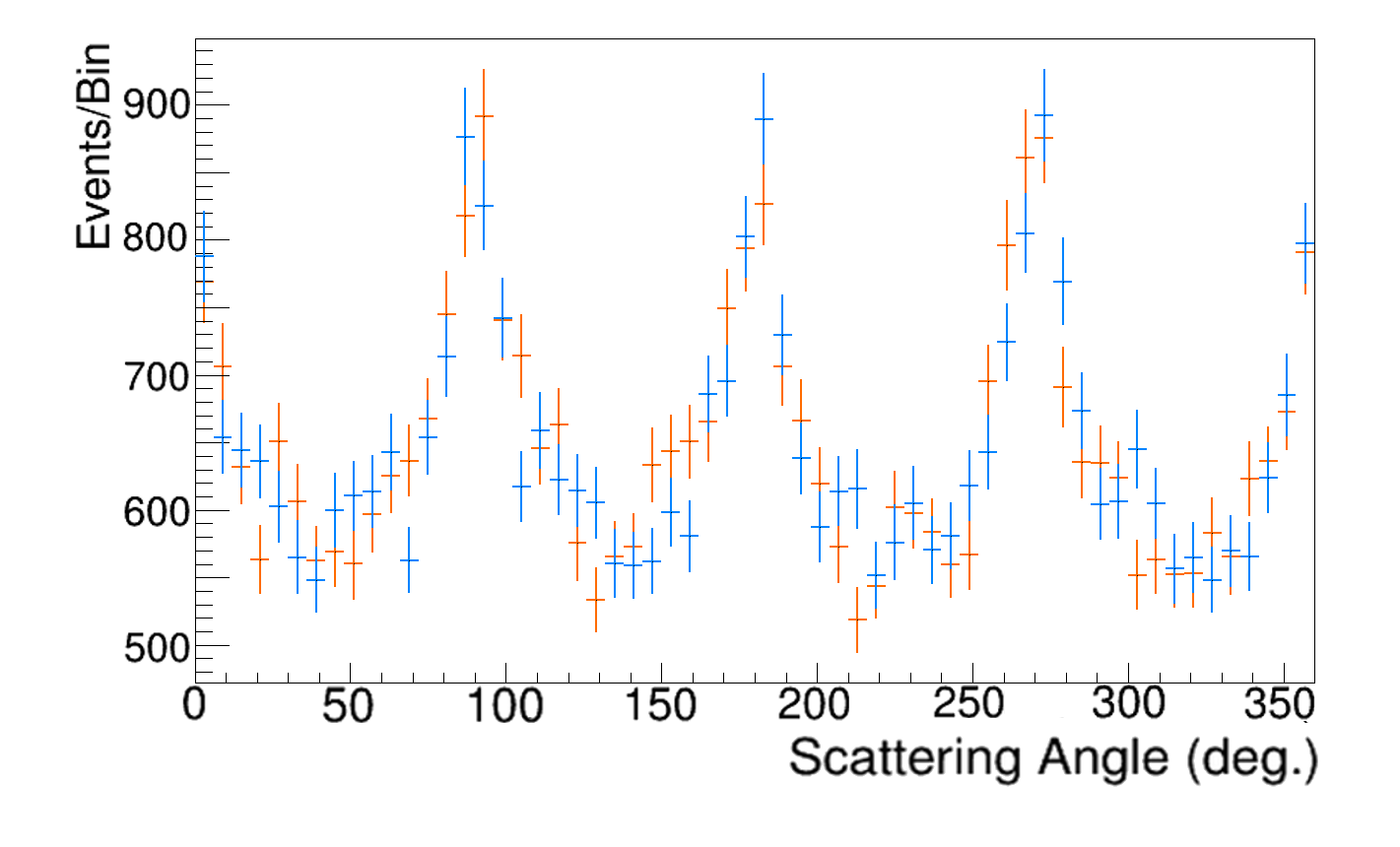}
  \caption{\textbf{The unnormalized modulation curves produced from two different background time intervals around GRB 170114A.} The orange crosses are produced when selecting a gap interval of 5 seconds between the GRB and the background whereas the blue are produced when this interval is set to 90 seconds. The statistical uncertainties are displayed for both histograms.}
\label{bg_2}
\end{figure}

The analysis for GRB 170114A is furthermore repeated using three different background time intervals all with the same length, taken to be 10 times longer than the signal time interval. First only a background  prior to the burst is selected. Second, only a background interval after the burst was selected and finally a background time interval half of which is before and half of which is after the GRB. The resulting best fitting polarization degrees and angles are presented in Supplementary Table~\ref{tab:post_prior_bg} together with the minimum $\chi^2$/NDF value found for each analysis.

\begin {table}[!ht]
\small
\begin {center}
\caption{\textbf{The resulting PD and PA values together with the corresponding $\chi^2$/NDF acquired when selecting different background time intervals in the analysis.}}
\begin{tabular}{| l | c | c | c |}
  \hline			
  time interval & PD. (\%) & PA (deg.) & minimum $\chi^2$/NDF \\
  \hline			

  prior & 5\% & $164^\circ$ & 1.15 \\
  post & 7\% & $164^\circ$ & 1.27 \\
  post and prior & 4\% & $164^\circ$ & 1.21 \\
\hline
\end{tabular}
\label{tab:post_prior_bg}
\end{center}
\end{table}

It can be concluded from this study that the background in POLAR is sufficiently stable for the background selection procedure to have little influence on the final result.

\subsection{Spectral influence}

In order to study the influence of the input spectrum used in the simulations on the analysis results, GRB 170206A is studied in more detail. The energy spectrum of GRB 170206A was both reported by Fermi-GBM and by Konus-Wind ~\cite{GCN_170206A}. The measurement result presented by both instruments differ as can be seen in the value of $\alpha$ reported by Fermi-GBM which is $-0.28\pm0.04$ whereas that by Konus-Wind is $0.00\pm0.2$. The GRB is therefore analyzed both using the best fit spectrum provided by Fermi-GBM and the spectrum provided by Konus-Wind. The analysis results for PD and PA, along with the $\chi^2/$NDF from both the analyses are provided in Supplementary Table \ref{tab:spec}. We can conclude that the different spectra do not significantly influence the final results.

\begin{table}[!ht]
\centering
\caption{\textbf{The resulting PD and PA values together with the corresponding $\chi^2$/NDF acquired when using spectral parameters from different instruments.}}
\begin{tabular}{| l | c | c | c | }
  \hline			
  spectrum source & PD (\%) & PA (deg.) & $\chi^2$/NDF \\
  \hline			
  Fermi-GBM & 10\% & $106^\circ$ & 26.3/27 \\
  Konus-Wind & 9\% & $106^\circ$ & 26.5/27 \\
  \hline
\end{tabular}
\label{tab:spec}
\end{table}

\subsection{GRB location influence}

In order to study the influence on the final analysis results for PD and PA of the position of the GRB with respect to POLAR, the analysis of GRB 170114A is performed in more detail. Although the reported measurement uncertainties in the GRB location are taken into account in the systematic errors of the polarization analysis, for a GRB such as 170114A the location is based on that provided by a single instrument, here Fermi-GBM~\cite{GCN_170114A}. The reported uncertainties in the GCN do not include systematic uncertainties which are of the order of $3.7^\circ$ ref.~\cite{Connaughton2015}. For GRB 170114A the influence of a significant error on the location of the GRB is therefore studied by changing the incoming angle in the simulations by as much as $3.7^\circ$. Two separate analyses are performed, one where $3.7^\circ$ was added to the $\theta$ angle (the off-axis angle), and one to the $\phi$ angle (within the detector plane). The influence on the measured PD are found to be minimal, as shown in Supplementary Table \ref{tab:location}, while the influence on the measured PA is around $10^\circ$. However, as PD is low and thus PA is poorly constrained for this GRB, we can therefore conclude that even if there is a significant systematic error in the provided location of the GRB it would not alter our results for this GRB significantly.

\begin{table}[!ht]
\centering
\caption{\textbf{The resulting PD and PA values together with the corresponding $\chi^2$/NDF acquired under different incident angle conditions.}}
\begin{tabular}{| l | c | c | c |}
  \hline			
  angle & PD (\%) & PA (deg.) & $\chi^2$/NDF \\
  \hline			
  normal & 4\% & $164^\circ$ & 32.3/27\\
  $\theta$ + 3.7$^\circ$ & 3\% & $154^\circ$ & 33.5/27\\
  $\phi$ + 3.7$^\circ$ & 4\% & $162^\circ$ & 34.4/27 \\
  \hline
\end{tabular}
\label{tab:location}
\end{table}

\section{Time-integrated polarization analysis}\label{sec:time_int}

The five GRBs are analyzed using the method described in the methods section of the main paper. Here the following parameters used for the analysis will be presented for each of the GRBs as listed in Supplementary Table~\ref{tab:grb_pars}:

\begin{itemize}
 \item Selected start time of the signal (UT), tagged as T$_{\rm start}$
 \item Selected stop time of the signal (UT), tagged as T$_{\rm stop}$
 \item Time interval between signal and background, tagged as T$_{\rm interval}$
 \item Ratio of background over signal time interval, tagged as R$_{\rm BOS}$
 \item The best fit GRB spectra model, tagged as Model
 \item The referenced instruments where the GRB spectra come from, tagged as Instrument (Spectrum).
 \item Value of $\alpha$ for the Band function as used in the simulations
 \item Value of $\beta$ for the Band function as used in the simulations
 \item Value of $E_{\rm peak}$ for the Band function as used in the simulations
 \item The referenced instruments where the GRB localization come from, tagged as Instrument (Localization).
 \item Incoming $\theta$ angle (angle from zenith) as used in the simulations
 \item Incoming $\phi$ angle (angle in the detector plane) as used in the simulations
 \item The error of localization, tagged Error (Localization).
\end{itemize}

Furthermore the following results from the analysis will be provided for each GRB as listed in Supplementary Table~\ref{tab:grb_results}:

\begin{itemize}
 \item Number of signal events (after background subtraction), tagged as $Events_{\rm S}$
 \item Number of background events (after normalization to signal time interval), tagged as $Events_{\rm B}$
 \item $M_{100}$ value based on simulations
 \item The Minimum Detectable Polarization (MDP), as defined in~\cite{Weisskopf}, for this GRB based on the above three parameters
 \item Best fit results in PD and PA (with PA in galactic coordinates)
 \item The $\chi^2/$NDF value of the result
 \item The probability for the result to be produced by a flux with a PD<2\%, as calculated using the Bayesian method.
 \item Upper limit for polarization degree (99\% credible), tagged as PD$_{\rm up}$ as calculated using the Bayesian method.
\end{itemize}

\begin {table}[!ht]
\footnotesize
\begin {center}
\caption{\textbf{List of GRB parameters used for the polarization analysis.} For all except one a Band function was used to model the spectrum, for GRB 170127C a single cut-off power law (*CPL) model was used.}
\begin{tabular}{ | p{2.2cm} | p{2.1cm} | p{2.1cm} | p{2.1cm} | p{2.1cm} | p{2.1cm} |}  \hline
GRB & 161218A ref.~\cite{GCN_161218A} & 170101A ref.~\cite{GCN_170101A} & 170114A ref.~\cite{GCN_170114A} & 170127C ref.~\cite{GCN_170127C} & 170206A ref.~\cite{GCN_170206A} \\ \hline\hline
T$_{\rm start}$ & 2016-12-18 T03:47:34.0 & 2017-01-01 T02:26:00.66 & 2017-01-14 T22:01:09 & 2017-01-27 T01:35:47.75 & 2017-02-06 T10:51:57.400 \\ \hline
T$_{\rm stop}$ & 2016-12-18 T03:47:40.5 & 2017-01-01 T02:26:02.16 & 2017-01-14 T22:01:18 & 2017-01-27 T01:35:47.95 & 2017-02-06 T10:51:59.400 \\ \hline
T$_{\rm interval}$ & 5 seconds & 5 seconds & 5 seconds & 90 seconds & 5 seconds \\ \hline
R$_{\rm BOS}$ & 14 & 20 & 10 & 50 & 20 \\ \hline
Instrument (Spectrum) & Konus-Wind & Konus-Wind & Fermi-GBM & Fermi-GBM & Fermi-GBM \\ \hline
Model & Band & Band & Band & CPL* & Band \\ \hline
$\alpha$ & -0.28${_{-0.21}^{+0.25}}$ & -1.44${_{-0.13}^{+0.17}}$ & -0.84$\pm$0.04 & -0.27$\pm$0.09 & -0.28$\pm$0.04 \\ \hline
$\beta$ & -3.40${_{-1.17}^{+0.43}}$ & -2.49${_{-0.65}^{+0.23}}$ & -1.99$\pm$0.06 & N/A & -2.55$\pm$0.12 \\ \hline
$E_{\rm peak}$ (keV) & 128${_{-8}^{+8}}$ & 123${_{-21}^{+23}}$ & 237$\pm$19 & 859$\pm$34 & 341$\pm$13  \\ \hline
Instrument (Localization) & Swift-BAT & Swift-BAT & Fermi-GBM & Fermi-LAT & IPN \\ \hline
$\theta$ & $24.3^\circ$ & $5.6^\circ$ & $26.5^\circ$ & $41.8^\circ$ & $18.3^\circ$ \\ \hline
$\phi$ & $356.1^\circ$ & $68.9^\circ$ & $5.4^\circ$ & $157.6^\circ$ & $147.2^\circ$ \\ \hline
Error $\qquad$ (Localization) & 3$'$ (90\%) & 1.1$'$ (90\%) & 1$^\circ$ (1$\sigma$) & 0.4$^\circ$ (90\%) & 1156 $(')^2$ (3$\sigma$) \\ \hline
\end{tabular}
\label{tab:grb_pars}
\end{center}
\end {table}

\begin {table}[!ht]
\begin {center}
\caption{\textbf{List of results from the polarization analysis.}}
\begin{tabular}{ | p{3.1cm} | p{1.6cm} | p{1.6cm} | p{1.6cm} | p{1.6cm} | p{1.6cm} |}  \hline
GRB & 161218A & 170101A & 170114A & 170127C & 170206A \\ \hline\hline
Events$_{\rm S}$ & 2079 & 1958 & 3709 & 685 & 2566 \\ \hline
Events$_{\rm B}$ & 3876 & 709 & 3971 & 98 & 796 \\ \hline
M$_{100}$ & 37\% & 38.5\% & 37.0\% & 30.0\% & 37.5\%  \\ \hline
MDP & 43.0\% & 29.4\% & 27.4\% & 58.4\% & 25.9\% \\ \hline
PD & 9\% &  8\% & 4\% & 11\% & 10\% \\ \hline
PA & $40^\circ$ & $164^\circ$ & $164^\circ$ & $38^\circ$ & $106^\circ$  \\ \hline
$\chi^2$/NDF & 20.6/27 & 33.7/27 & 32.3/27 & 5.3/12 & 26.3/27 \\ \hline
Prob(PD$<$2\%) & 9\% & 13\% & 5.8\% & 12\% & 14\% \\ \hline
PD$_{\rm up}$ (99\%) & 45\% & 31\% & 28\% & 67\% & 31\% \\ \hline
\end{tabular}
\label{tab:grb_results}
\end{center}
\end {table}

\clearpage
\section{Time Dependent Analysis Results}

\begin{figure}[!ht]
  \centering
    \includegraphics[width=12 cm]{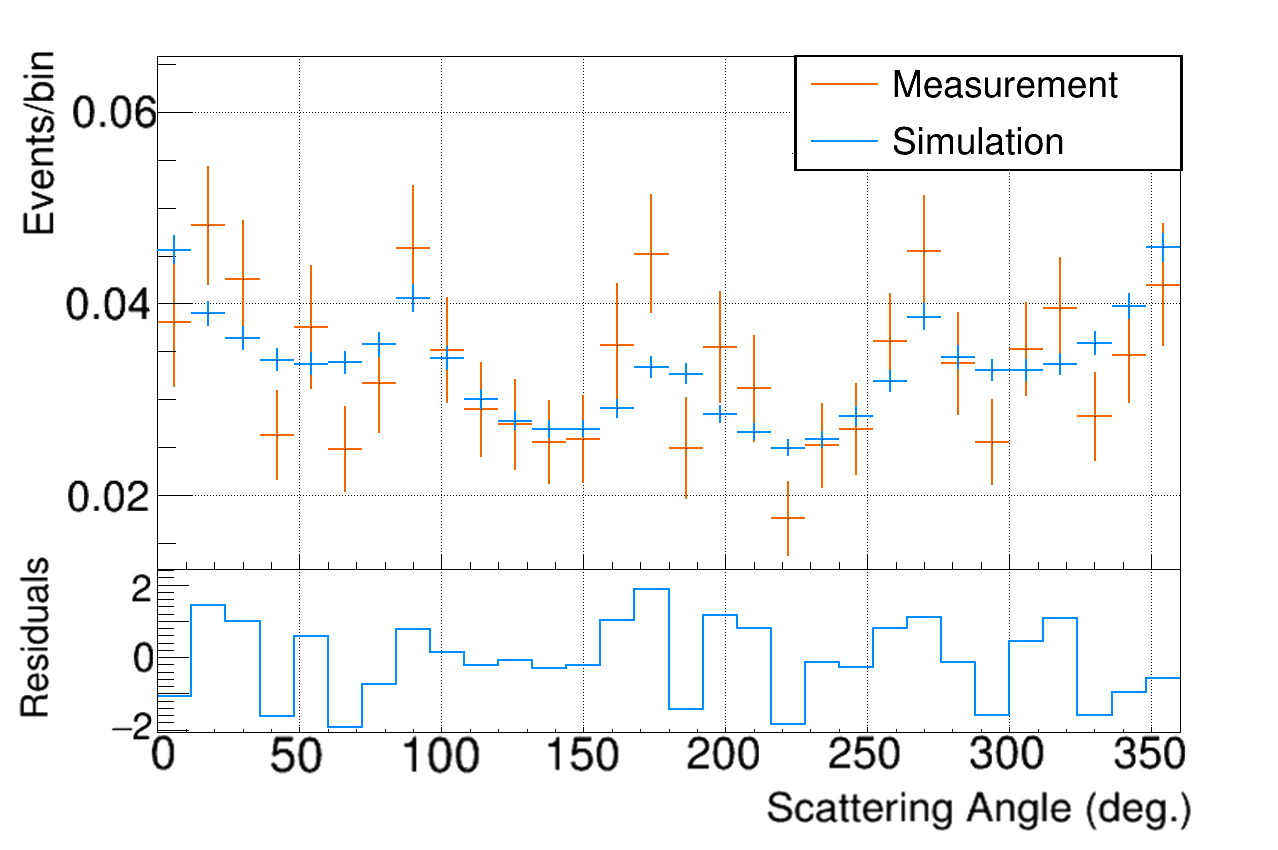}
  \caption[Modulation curves for 170114A for time bin 1]
 {\textbf{Modulation curves for 170114A for the first time bin.} The measured modulation curve is shown in orange together with the best fitting simulated modulation curve, in blue, for the first time bin of GRB 170114A. The residuals, which result in a $\chi^2/$NDF of 34.1/27, for the two histograms are shown in the histogram below. The uncertainties displayed for the measured histogram include the statistical uncertainty, for the simulated histogram both the systematic and statistical uncertainties are displayed.}
\label{mod_170114A_p1}
\end{figure}

\begin{figure}[!ht]
  \centering
    \includegraphics[width=11 cm]{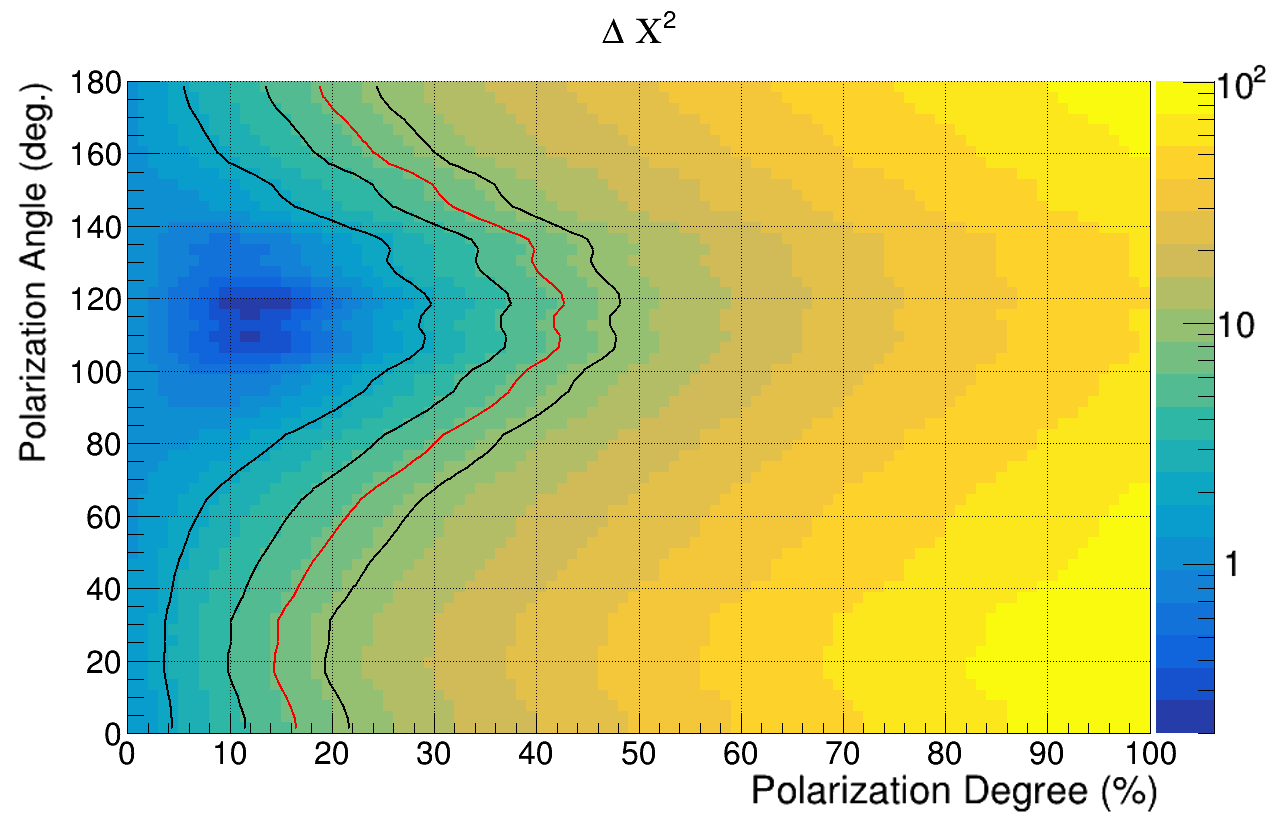}
  \caption[Likelihood plot Geneva for 170114A for time bin 1]
{\textbf{$\Delta\chi^2$ plot for the first 3 seconds of 170114A resulting from the analysis of the individual time bin.} Different likelihood intervals are drawn, from inner to outer these are the 68\%, 90\% confidence contours for 2 d.o.f., followed by the 99\% confidence contour for 1 d.o.f. (in red) and lastly the 99\% confidence contour for 2 d.o.f.}
\label{DChi_170114A_p1}
\end{figure}

\begin{figure}[!ht]
  \centering
    \includegraphics[width=12 cm]{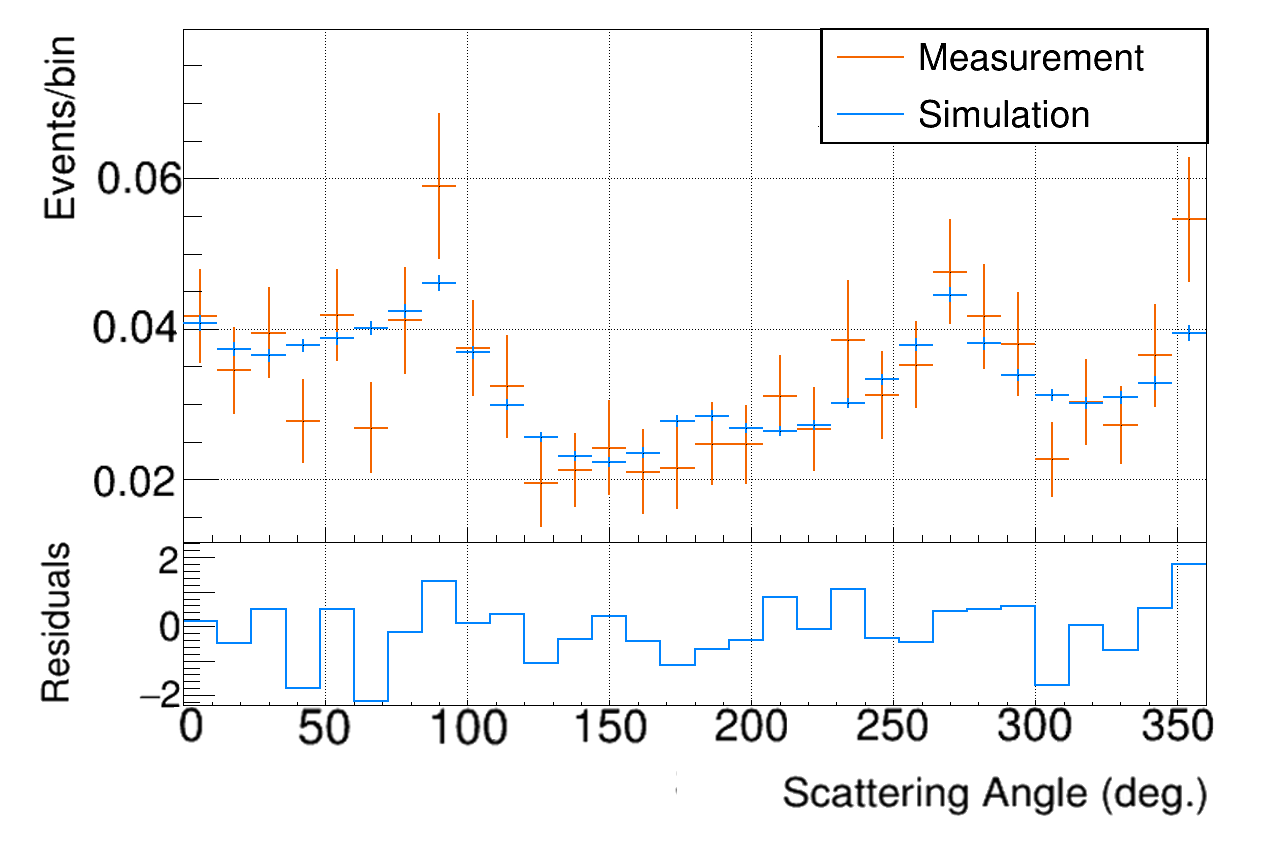}
  \caption[Modulation curves for 170114A for time bin 2]
 {\textbf{Modulation curves for 170114A for the second time bin.} The measured modulation curve is shown in orange together with the best best fitting simulated modulation curve, in blue, for the second time bin of GRB 170114A. The residuals, which result in a $\chi^2/$NDF of 23.9/27, for the two histograms are shown in the histogram below. The uncertainties displayed for the measured histogram include the statistical uncertainty, for the simulated histogram both the systematic and statistical uncertainties are displayed.}
\label{mod_170114A_p2}
\end{figure}

\begin{figure}[!ht]
  \centering
    \includegraphics[width=12 cm]{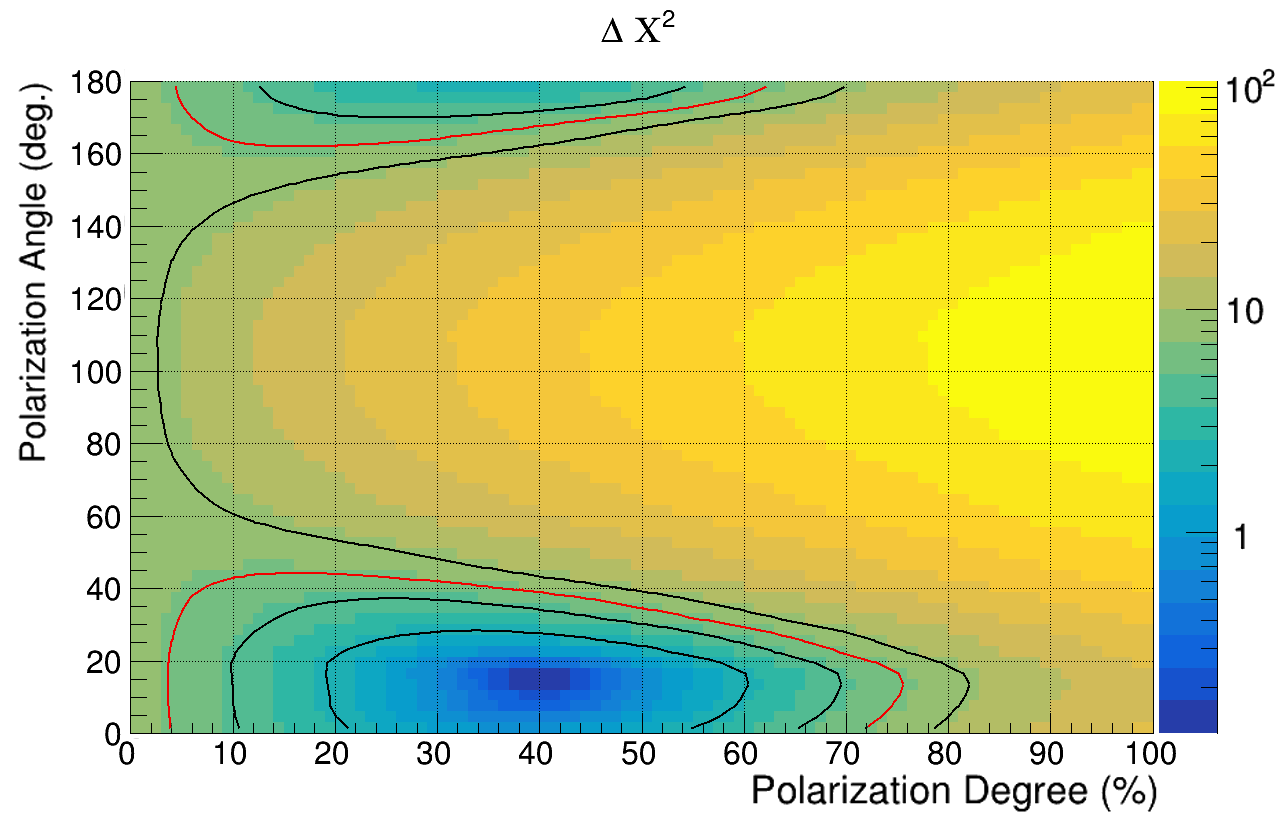}
  \caption[Likelihood plot Geneva for 170114A for time bin 2]
{\textbf{The $\Delta\chi^2$ distribution for the second time bin of GRB 170114A.} The $\Delta\chi^2$ values are shown for the second 3 second long time interval of 170114A resulting from the analysis of the individual time bin. Likelihood intervals are additionally drawn, from inner to outer these are the 68\%, 90\% confidence contours for 2 d.o.f., followed by the 99\% confidence contour for 1 d.o.f. (in red) and lastly the 99\% confidence contour for 2 d.o.f.}
\label{DChi_170114A_p2}
\end{figure}

\begin{figure}[!ht]
  \centering
    \includegraphics[width=12 cm]{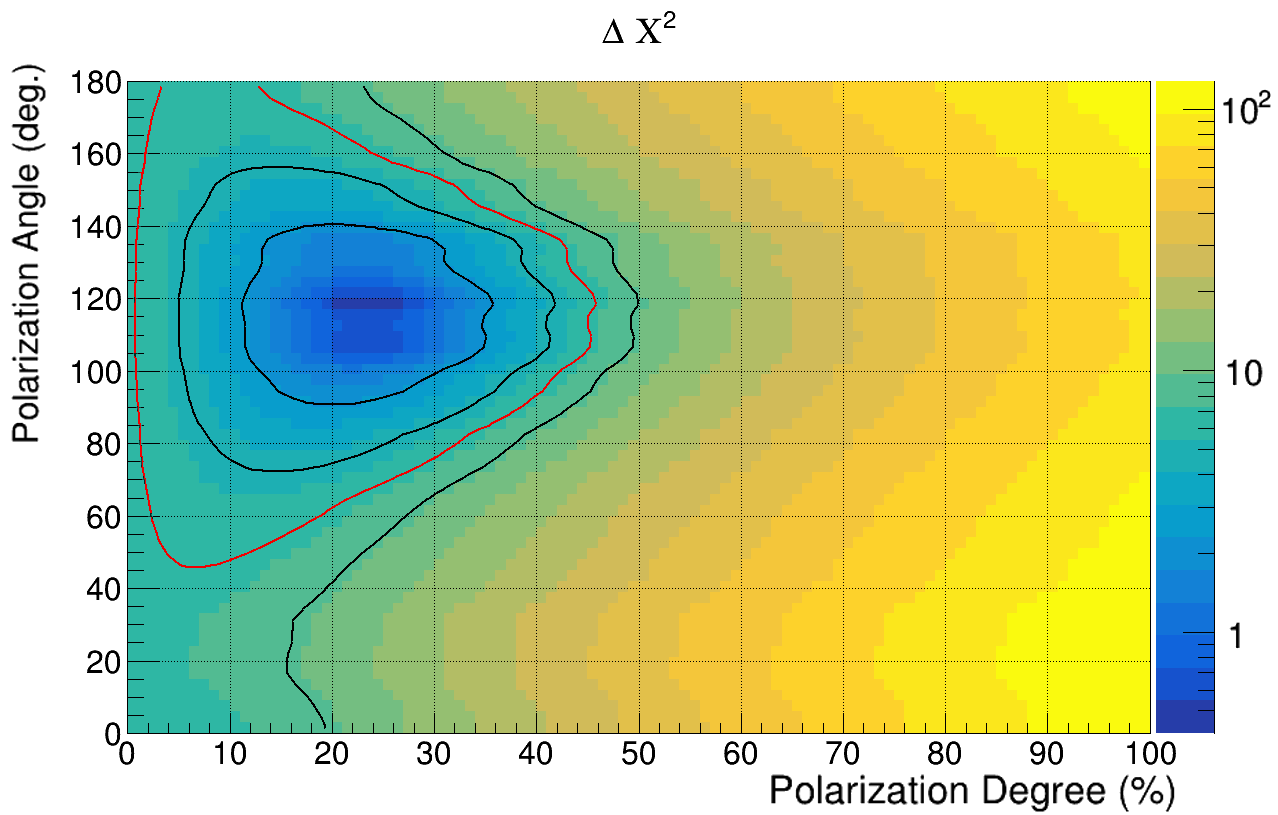}
  \caption[Likelihood for the combined fit of the 2 time bins of 170114A]
{\textbf{The $\Delta\chi^2$ distribution resulting from a combined time resolved analysis using 2 time intervals.} The $\Delta\chi^2$  distribution, shown here for the first time bin, is produced using the combined analysis of the two time bins of 170114A. From inner to outer the 68\%, 90\% confidence contours are drawn for 2 d.o.f., followed by the 99\% confidence contour for 1 d.o.f. (in red) and lastly the 99\% confidence contour for 2 d.o.f.}
\label{comb_170114A}
\end{figure}

\begin{figure}[!ht]
  \centering
    \includegraphics[width=12 cm]{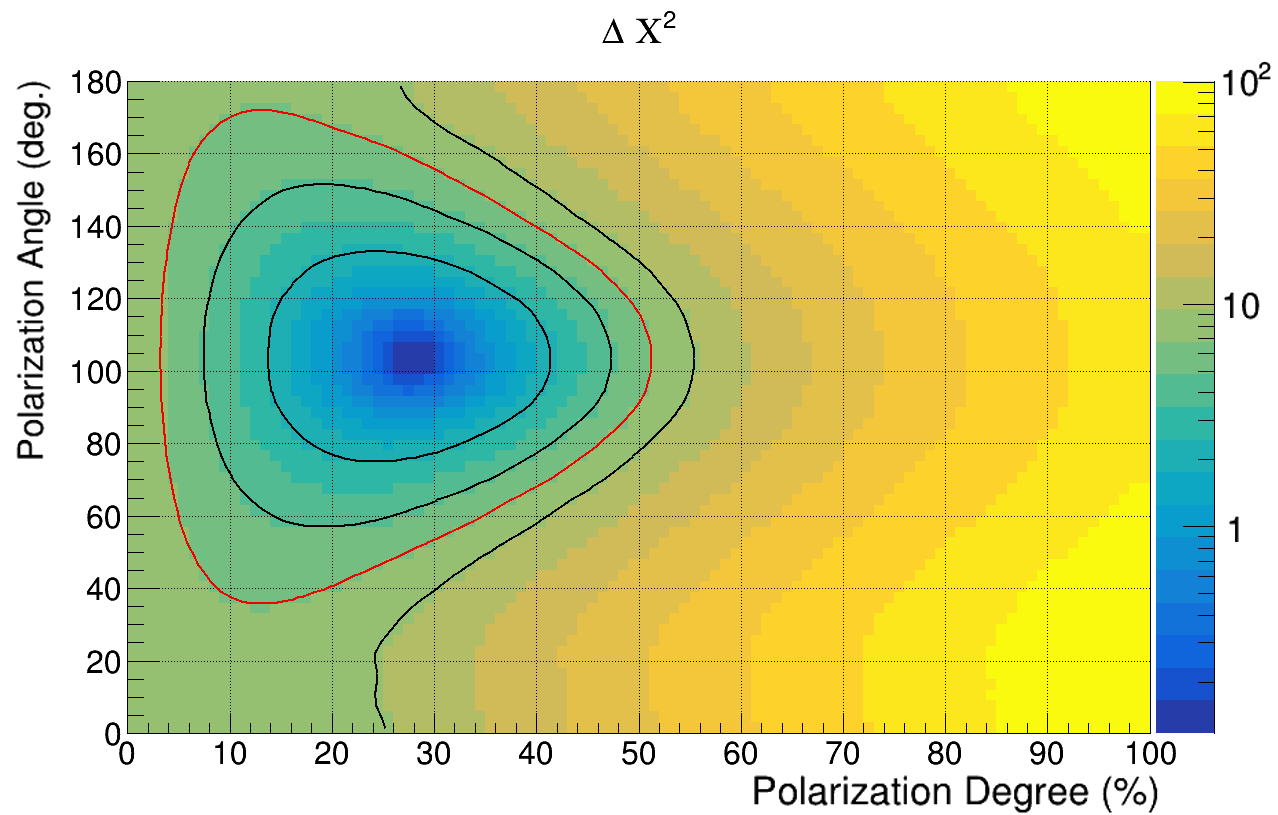}
  \caption[Likelihood for the combined fit of the 2 time bins of 170114A]
{\textbf{The $\Delta\chi^2$ distribution resulting from a combined time resolved analysis using 3 time intervals.} The $\Delta\chi^2$ distribution, shown here for the first time bin, is produced using the combined analysis of the three time bins of 170114A. From inner to outer the 68\%, 90\% confidence contours are drawn for 2 d.o.f., followed by the 99\% confidence contour for 1 d.o.f. (in red) and lastly the 99\% confidence contour for 2 d.o.f. Similar to Supplementary Figure \ref{comb_170114A} this figure results from performing the combined fit on time bins of GRB170114A, however here the data was divided into 3 time bins. }
\label{comb_170114A_3}
\end{figure}



\begin{thebibliography}{99}

\bibitem{Woosley1993} Woosley, S. E. Gamma-ray bursts from stellar mass accretion disks around black holes. \textit{Astrophys. J.} \textbf{405}, 273-277 (1993).
\bibitem{Iwamoto} Iwamoto, K. et al. A hypernova model for the supernova associated with the {$\gamma$}-ray burst of 25 April 1998. \textit{Nature.} \textbf{395}, 672-674 (1998).
\bibitem{MacFayden1999} MacFadyen, A. I. \& Woosley, S. E. Collapsars: Gamma-Ray Bursts and Explosions in "Failed Supernovae", \textit{Astrophys. J.} \textbf{524},  262-289 (1999).
\bibitem{Gehrels2013} Gehrels, N. \& Razzaque, S. Gamma-ray bursts in the swift-Fermi era. \textit{Front. of Phys.} \textbf{8}, 661-678 (2013).
\bibitem{Toma2009} Toma, K. et al. Statistical Properties of Gamma-Ray Burst Polarization. \textit{Astrophys. J.} \textbf{698}, 1042-1054 (2009).
\bibitem{Covino:2016cuw} Covino, S. \& G\"{o}tz, D. Polarization of prompt and afterglow emission of Gamma-Ray Bursts. \textit{Astron. Astrophys. Trans.} \textbf{29}, 205-239 (2016).
\bibitem{MCCONNELL20171} McConnel, M. High energy polarimetry of prompt GRB emission. \textit{New Astr. Rev.} \textbf{76}, 1-21 (2017).
\bibitem{COSI2} Lowell, A.W. et al. Polarimetric Analysis of the Long Duration Gamma-Ray Burst GRB 160530A With the Balloon Borne Compton Spectrometer and Imager. \textit{Astrophys. J. L.} \textbf{848}, 119-129 (2017).
\bibitem{GAPGRB2} Yonetoku, D. et al. Magnetic Structures in Gamma-Ray Burst Jets Probed by Gamma-Ray Polarization. \textit{Astrophys. J. L.} \textbf{758}, L1 (2012).
\bibitem{GAP} Yonetoku, D. et al. Detection of Gamma-Ray Polarization in Prompt Emission of GRB 100826A. \textit{Astrophys. J. L.} \textbf{743}, L30 (2011).
\bibitem{Produit2017} Produit, N. et al. Design and construction of the POLAR detector. \textit{Nucl. Instrum. Meth. Phys. Res. A.} \textbf{877}, 259-268 (2018).
\bibitem{LI2018} Li, Z. H. et al. In-orbit instrument performance study and calibration for POLAR polarization measurements. \textit{Nucl. Instrum. Meth. Phys. Res. A.} \textbf{900}, 8-24 (2018).
\bibitem{Kole2017} Kole, M. et al. Instrument performance and simulation verification of the POLAR detector. \textit{Nucl. Instrum. Meth. Phys. Res. A.} \textbf{872}, 28-40 (2017).
\bibitem{1995AAS...186.5301K} Koshut, T. M. et al. T$_{90}$ as a Measurment of the Duration of GRBs. \textit{Am. Astr. Soc. M. Abstr.} \textbf{27}, 886 (1995).
\bibitem{Maier2014} Maier, D., Tenzer, C. \& Santangelo, A. Point and Interval Estimation on the Degree and the Angle of Polarization: A Bayesian Approach. \textit{Publ. Astro. Soc. Pacific.} \textbf{126}, 459-468 (2014).
\bibitem{KCWestfold1959} Westfold, K. C. The Polarization of Synchrotron Radiation. \textit{Astrophys. J.} \textbf{130}, 241-258 (1959).
\bibitem{WHMcMaster1961} McMaster, W. H. Matrix Representation of Polarization. \textit{Rev. Mod. Phys.} \textbf{33}, 8-28 (1961).
\bibitem{Lazzati2004} Lazzati, D., Rossi, E. Ghisellini, G. \& Rees, M.J. Compton drag as a mechanism for very high linear polarization in gamma-ray bursts. \textit{Mon. Not. R. Astron. Soc.} \textbf{347}, L1-L5 (2004).
\bibitem{JG2003} Granot, J. The Most Probable Cause for the High Gamma-Ray Polarization in GRB 021206. \textit{Astrophys. J. L.} \textbf{596}, L17-L21 (2003).
\bibitem{MXLan2018} Lan, M.X., Wu, X.F. \& Dai, Z. G. Gamma-Ray Burst Optical Afterglows with Two-Component Jets: Polarization Evolution Revisited, \textit{Astrophys. J.} \textbf{860}, 44-51 (2018).

\bibitem{Rossi2004} Rossi, E. M., Lazzati, D., Salmonson, J.D. \& Ghisellini G. The polarization of afterglow emission reveals gamma-ray bursts jet structure \textit{Mon. Not. R. Astron. Soc.} \textbf{354}, 86-100 (2004).

\bibitem{Gill2018} Gill, R. \& Granot, J. Afterglow Imaging and Polarization of Misaligned Structured GRB Jets and Cocoons: Breaking the Degeneracy in GRB 170817A. \textit{Mon. Not. Roy. Astron. Soc.} \textbf{478}, 4128-4141 (2018).

\bibitem{MJRees1994} Rees, M. J. \& Meszaros, P. Unsteady outflow models for cosmological gamma-ray bursts. \textit{Astrophys. J. L.} \textbf{430}, L93-L96 (1994).
\bibitem{BP1994} Paczynski, B. \& Xu, G. Neutrino bursts from gamma-ray bursts. \textit{Astrophys. J.} \textbf{427}, 708-713 (1994).
\bibitem{PMeszaros2000} M\'{e}sz\'{a}ros, P. \& Rees, M. J. Steep Slopes and Preferred Breaks in Gamma-Ray Burst Spectra: The Role of Photospheres and Comptonization. \textit{The Astrophys. J.} \textbf{530}, 292-298 (2000).
\bibitem{MJRees2005} Rees, M. J. \& M\'{e}sz\'{a}ros, P. Dissipative Photosphere Models of Gamma-Ray Bursts and X-Ray Flashes. \textit{The Astrophys. J.} \textbf{628}, 847-852 (2005).
\bibitem{BingZhang2011} Zhang, B. \& Yan, H. The Internal-collision-induced Magnetic Reconnection and Turbulence (ICMART) Model of Gamma-ray Bursts. \textit{Astrophys. J.} \textbf{726}, 90-113 (2011).
\bibitem{Christoffer2016} Lundman, C., Vurm, I. \& Beloborodov, A. M. Polarization of gamma-ray bursts in the dissipative photosphere model. \textit{Astrophys. J.} \textbf{856} (2018).
\bibitem{WeiDeng2016} Deng, W., Zhang, H., Zhang, B. \& Li, H. Collision-induced Magnetic Reconnection and a Unified Interpretation of Polarization Properties of GRBs and Blazars. \textit{Astrophys. J. L.} \textbf{821}, L12-L19 (2016).

\setcounter{firstbib}{\value{enumiv}}

\end{thebibliography}

\begin{thebibliography}{9}

\setcounter{enumiv}{\value{firstbib}}

\bibitem{Xiong2017} Xiong, S. L. et al. Overview of the GRB observation by POLAR. \textit{PoS (ICRC 2017).} \textbf{640}, (2017).
\bibitem{light_curves} POLAR Light Curves. \url{https://www.isdc.unige.ch/polar/lc/}, accessed 2-July-2018
\bibitem{Band1993} Band, D. et al. BATSE observations of gamma-ray burst spectra. I - Spectral diversity. \textit{Astrophys. J.} \textbf{413}, 281-292 (1993).
\bibitem{Avni1976}  Avni, Y. Energy spectra of X-ray clusters of galaxies. \textit{Astrophys. J.} \textbf{210}, 642-646 (1976).
\bibitem{GCN_170114A} GCN for GRB 170114A. \url{https://gcn.gsfc.nasa.gov/other/170114A.gcn3}, accessed 13-June-2018.


\end{thebibliography}

\begin{thebibliography}{99}

\bibitem{Covino:2016cuw} Covino, S. \& G\"{o}tz, D. Polarization of prompt and afterglow emission of Gamma-Ray Bursts. \textit{Astron. Astrophys. Trans.} \textbf{29}, 205-239 (2016).

\bibitem{COSI2} Lowell, A. W. et al. Polarimetric Analysis of the Long Duration Gamma-Ray Burst GRB 160530A With the Balloon Borne Compton Spectrometer and Imager. \textit{ Astrophys. J.}. \textbf{848}， 119 (2017).
	
\bibitem{GAPGRB2} Yonetoku, D. et al. Magnetic Structures in Gamma-Ray Burst Jets Probed by Gamma-Ray Polarization. \textit{Astrophys. J. L.} \textbf{758}, L1 (2012).
\bibitem{GAP} Yonetoku, D. et al. Detection of Gamma-Ray Polarization in Prompt Emission of GRB 100826A. \textit{Astrophys. J. L.} \textbf{743}, L30 (2011).
    
\bibitem{RHESSI2} Coburn, W. \& Boggs, S. E. RHESSI search for polarization in prompt GRB emission. \textit{AAS/High Energy Astrophys. Div. \#7.} \textbf{35}, 623 (2003).

\bibitem{RHESSI1} Wigger, C., Hajdas, W., Arzner, K., {G{\"u}del}, M., \& Zehnder, A. Polarization from GRB021206: No constraints from reanalysis of RHESSI data. \textit{Nuovo Cimento C Geophys. Space Phys. C.} \textbf{28}, 265 (2005).

\bibitem{IBIS1} G\"{o}tz, D. et al. GRB 140206A: the most distant polarized gamma-ray burst. \textit{Mon. Not. R. Astron. Soc.}\textbf{444}, 2776-2782 (2014).

\bibitem{IBIS3} G\"{o}tz, D., Covino, S., Fern\'{a}ndez-Soto, A., Laurent, P. \& Bo\v{s}njak, \v{Z}. The polarized gamma-ray burst GRB 061122. \textit{Mon. Not. R. Astron. Soc.} \textbf{431} 3550-3556 (2013).

\bibitem{IBIS2} G\"{o}tz, D., Laurent, P.,Lebrun, F., Daigne, F. \& Bo\v{s}njak, \v{Z}. Variable Polarization Measured in the Prompt Emission of GRB 041219A Using IBIS on Board INTEGRAL. \textit{Astrophy. J. L.} \textbf{695}, L208 (2009).

\bibitem{SPI} Kalemci, E., Boggs, S. E., Kouveliotou, C., Finger, M. \& Baring, M. G. Search for Polarization from the Prompt Gamma-Ray Emission of GRB 041219a with SPI on INTEGRAL. \textit{Astrophys. J. Suppl. S.} \textbf{169}, 75 (2007).

\bibitem{BATSE1} Wills, D. R. et al., Evidence of polarisation in the prompt gamma-ray emission from GRB 930131 and GRB 960924. \textit{Astron. Astrophys.} \textbf{439}, 245-253 (2005).

\bibitem{LI2018} Li, Z. H. et al. In-orbit instrument performance study and calibration for POLAR polarization measurements. \textit{Nucl. Instrum. Meth. Phys. Res. A.} \textbf{900}, 8-24 (2018).

\bibitem{KONUS_GCN} Konus-Wind Light Curves for GRB 170127C. \url{http://www.ioffe.ru/LEA/GRBs/GRB170127_T05744/}, accessed 13-June-2018.

\bibitem{GCN_170206A} GCN for GRB 170206A. \url{https://gcn.gsfc.nasa.gov/other/170206A.gcn3}, accessed 13-June-2018.

\bibitem{GCN_170114A} GCN for GRB 170114A. \url{https://gcn.gsfc.nasa.gov/other/170114A.gcn3}, accessed 13-June-2018.

\bibitem{Connaughton2015} Connaughton, V. et al. Localization of Gamma-Ray Bursts Using the Fermi Gamma-Ray Burst Monitor. \textit{Astrophys. J. Suppl. L.} \textbf{216}, 32 (2015).

\bibitem{Weisskopf} Weisskopf, M. C., Elsner, R. F. \& O'Dell, S. L. On understanding the figures of merit for detection and measurement of x-ray polarization. \textit{Proc.SPIE.}. \textbf{7732}, 77320E (2010).

\bibitem{GCN_161218A} GCN for GRB 161218A. \url{https://gcn.gsfc.nasa.gov/other/161218A.gcn3}, accessed 13-June-2018.

\bibitem{GCN_170101A} GCN for GRB 170101A. \url{https://gcn.gsfc.nasa.gov/other/170101A.gcn3}, accessed 13-June-2018.

\bibitem{GCN_170127C} GCN for GRB 170127C. \url{https://gcn.gsfc.nasa.gov/other/170127C.gcn3}, accessed 13-June-2018.









\end{thebibliography}

\clearpage

\renewcommand{\refname}{References}

\end{document}